\documentclass[showpacs,amsmath,amssymb, prb,twocolumn]{revtex4}
\usepackage{graphicx}
\usepackage{dcolumn}
\usepackage{bm}

\begin{document}

\title{Interaction effects in non-equilibrium transport properties of a four-terminal topological corner junction}

\author{F. Romeo$^{1}$ and R. Citro$^{1,2}$}
\affiliation{$^{1}$Dipartimento di Fisica "E.R. Caianiello", Universit\`a  di Salerno, I-84084 Fisciano (SA), Italy\\
$^{2}$CNR-SPIN Salerno, I-84084 Fisciano (SA), Italy}

\date{\today}
\begin{abstract}
We study the transport properties of a four-terminal corner junction made by etching a two-dimensional topological insulator to form a quantum point contact (QPC). The QPC geometry enables inter-boundary tunneling processes allowing for the coupling among states with different helicity, while the tight confinement in the QPC region activates charging effects leading to the Coulomb blockade physics. Peculiar signatures of these effects are theoretically investigated using a scattering field theory modified to take into account the electron-electron interaction within a self-consistent mean-field approach. The current-voltage characteristics and the current fluctuations (noise) are derived beyond the linear response regime. Universal aspects of the thermal noise of the corner junction made of helical matter are also discussed.
\end{abstract}


\maketitle
\section{Introduction}
The recent discovery of topological insulators (TIs), both in three and in two dimensions, represents a promising advancement towards quantum technology of matter\cite{Hasan10, Qi11}. In fact, microscopic laws of physics are invariant under time reversal, but the transport of energy and information in real devices is an irreversible process. This irreversibility originates from the device-environment coupling and limits, for instance, the possibility of quantum computation. The TIs, bulk gapped materials exhibiting conducting channels at the boundaries\cite{day_phystoday_2008}, represent a possible technological platform to limit the decoherence induced by dissipative phenomena. Indeed, the peculiar electronic behavior predicted in devices based on topological materials, which can be designed as quantum spin-Hall systems\cite{kane_prl_2005}, could provide efficient spin injection protocols avoiding the complexity of working with nanoscale ferromagnetism\cite{bercioux2013}. Principles found in TIs could inspire the next-generation spintronic devices operating with low power consumption and/or performing topological protected quantum computation.
A specific example of two-dimensional TI is the HgTe/CdTe quantum well which has been demonstrated, both theoretically and experimentally, to have a single pair of helical edge states leading to a quantized conductance plateau when the Fermi energy lies in the bulk gap \cite{bernevig_science_2006, Konig07, roth_science_2009}. Quantized transport along the HgTe boundaries can be conveniently explained by an edge channel picture (Fig.~\ref{fig:fig1} (b)): Two states with opposite spin orientation propagate along opposite device edges in the same direction thus leading to a quantization\cite{Buttiker09} of the conductance in unit of $e^2/h$.\\
At the aim of exploiting topological protection and ballistic transport, several proposals of TI-based spin-transistor have been presented, being the working principles based on the sensitivity of the surface currents to the magnetization direction of a thin ferromagnetic barrier\cite{Kong11}, the Aharonov-Bohm and Fabry-P\'{e}rot interferometry \cite{qi_prb_2010,Dolcini11,Citro11,virtanen_prb_2011}, the field effect of a single gated HgTe nanoconstriction \cite{richter_prl_2011, chang_prb_2011,sternativo14}.\\
The simplest non-trivial TI-based device to probe a variety of transport properties exhibited by edge states can be obtained by creating a narrow constriction, i.e. a QPC, etched on a \textit{quantum spin hall insulator bar} as depicted in Fig.~\ref{fig:fig1} (a). Near-equilibrium spin and charge transport properties of similar two- and four-terminal geometries have been studied both in the non-interacting\cite{Citro11,Romeo12} and in the helical Luttinger liquid regime\cite{chamon_2009,johannesson_2009,schmidt_2011,Dolcetto12,ferraro13}. The emphasis of the previous studies was mainly on the conductance and the zero-frequency noise. Recently, finite frequency noise has been studied\cite{simon_2012,chung_2012} putting in evidence that tunneling processes of single- or two-particle contribute differently to the current fluctuations, the latter observation being important in discriminating multiple-quasiparticle (agglomerate) tunneling events in correlated systems.\\
However, in order to compare the theoretical models with the experiments performed on real devices one needs to take simultaneously into account non-equilibrium effects, non-perturbative inter-edge coupling, charging effects and Coulomb blockade physics, non-linear transport regimes. While most of these requirements are commonly met in free fermion models described by scattering approaches, the charging effects are completely missed. On the other hand, when interaction effects are taken into account inter-edge coupling is considered at perturbative level, while the effect of the interaction is modeled as uniformly distributed along the edge modes. The latter condition seems to be unlikely in real systems where charging effects are relevant only in the QPC region, i.e. where the tight confinement of the electronic densities enables unscreened Coulomb interaction effects. Furthermore, the perturbative nature (with respect to the inter-edge coupling) of the interacting theory limits its use to the linear response regime.\\
The aim of the present work is to give a comprehensive description of the interplay of all the above mentioned factors in the transport properties of a two-dimensional topological insulator by studying a minimal model of a four-terminal topological corner-junction. We give distinctive features of correlations and interference processes among helical edge states, useful for future experiments. In particular, our analysis, based on a modified scattering field theory, is focussed on a QPC geometry obtained by etching  a two-dimensional TI giving rise to a 4-terminal corner junction [Fig.~\ref{fig:fig1} (a)]. In this geometry inter-edge tunneling processes take place, while the tight confinement in the QPC region activates charging effects leading to a Coulomb-blockade-like physics. We describe all the peculiar signatures of these effects in the charge current and its fluctuation even beyond the linear response regime and suggest how to control edge-state transport by means of all electrical gating (side-gate control).\\
The organization of the paper is the following. In Sec.~\ref{sec:ham} we introduce an effective one-dimensional model to describe edge states at the surface of a two-dimensional topological insulator and their coupling in the QPC geometry. Here we present the model of four-terminal setup and its operational configuration. A discussion on the symmetry-preserving mean-field approximation of the interacting part of the Hamiltonian is also provided in Subsec.~\ref{ssec:spmfapp}. In Sec.~\ref{sec:scattering-field} we introduce the scattering field approach and the basic formalism used to calculate the current-voltage curves and the current-current correlations in terms of the scattering matrix elements, with emphasis on the role of the charging energy effect on the QPC. The modified boundary conditions and the self-consistency constraints are presented in Subsec.~\ref{ssec:bcs-self}. Results are presented in Sec.~\ref{sec:results}. In particular, in Subsec.~\ref{ssec:iv-diff-cond} we show the results concerning current-voltage characteristics and the differential conductance, with particular care in defining the properties of the Coulomb blockade regime. In Subsec.~\ref{ssec:cur-cur-noise} we introduce the general expression of the finite-frequency current-current correlation which also includes finite temperature effects, non-linear regime and non-equilibrium, non-local information and interaction effects. Thermal noise (\ref{ssec:thermal-noise}), local (\ref{ssec:transport-fluct}) and non-local (\ref{ssec:non-local-transport-fluct}) transport-induced fluctuations are carefully discussed and compared to the non-interacting case. Finally, we discuss our conclusions in Sec.~\ref{sec:concl}.

\section{Model Hamiltonian}
\label{sec:ham}
\begin{figure}[!t]
$\textbf{(a)}$ \includegraphics[clip,scale=0.3]{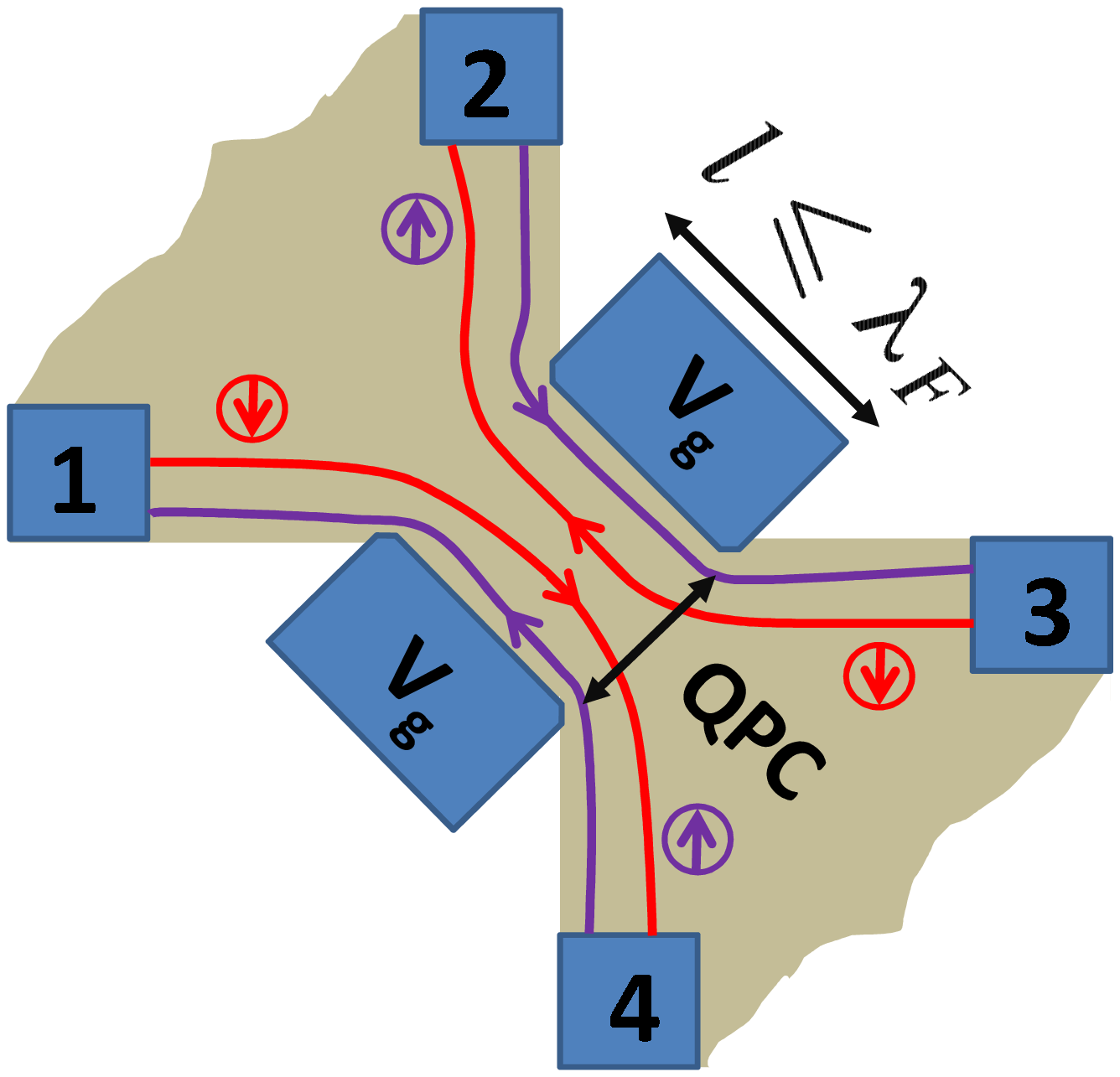}
$\textbf{(b)}$ \includegraphics[clip,scale=0.6]{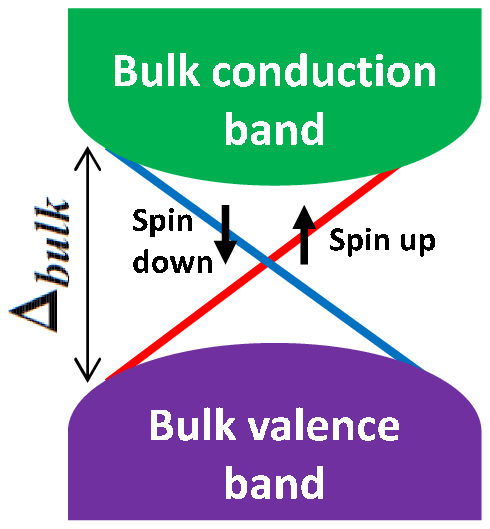}
\caption{(color online) $\textbf{(a)}$ Point-like constriction (i.e. linear dimension $l\leq \lambda_{F}$, $\lambda_{F}$ being the Fermi wave length) obtained by etching a two-dimensional topological insulator. A side-gate ($V_g$) can be used to tune the electronic density in the constriction, while a four-terminal geometry is considered. $\textbf{(b)}$ Sketch of the bands structure of a two-dimensional topological insulator. Helical states, having linear energy-momentum dispersion, describe the system as long as the particle energy ranges inside the bulk insulating gap $\Delta_{bulk}$. Coulomb interaction effects are assumed to be relevant only in the constriction region (QPC).}
\label{fig:fig1}
\end{figure}

In the following we formulate a minimal model describing the electronic transport in the corner junction depicted in Fig.\ref{fig:fig1} (a). In this system the confinement effects at the QPC locally activate the spin-orbit interaction and cause unscreened charging effects; while the former is responsible for spin-flipping tunneling effects, the latter gives origin to the Coulomb repulsion and possibly Coulomb blockade physics.
We thus consider an effective one-dimensional Hamiltonian $H$ describing edge states localized along the top and bottom boundaries of a quantum well defining a two-dimensional topological insulator and pinched to form a QPC:
\begin{equation}
\label{eq:hamiltonian}
H=H_0+H_{sp}+H_{sf}+H_C,
\end{equation}
where\cite{note1}:
\begin{eqnarray}
H_0=-i\hbar v_F \sum_{\sigma=\uparrow,\downarrow}\int dx && \lbrack : \psi^\dagger_{R\sigma}(x)\partial_x \psi_{R\sigma}(x): \nonumber \\
&&  -: \psi^\dagger_{L\bar{\sigma}}(x)\partial_x \psi_{L\bar{\sigma}}(x):\rbrack,
\end{eqnarray}
$\psi_{R(L)\sigma}$ represents the right (left) mover electron annihilation operator with spin $\sigma=\uparrow,\downarrow$, while $:\mathcal{O}:$ stands for the normal ordering of the operator $\mathcal{O}$ with respect to the equilibrium state defined by occupied energy levels below the Fermi sea. In our description, without loss of generality, we assume that spin-$\uparrow$ right movers $(R, \uparrow)$ and spin-$\downarrow$ left movers  $(L, \downarrow)$  flow along the top boundary while the spin-$\downarrow$ right movers $(R, \downarrow)$ and spin-$\uparrow$ left movers $(L, \uparrow)$ flow along the bottom boundary.

At the nanoconstriction ($x=0$) the edge modes are coupled by transverse confinement effects and the overlap between edge states belonging to different boundaries leads to the twofold effect, one of producing a local reactivation of the spin-orbit coupling\cite{so-react} and the other one of enabling Coulomb repulsion effects. Indeed, at the QPC ($x=0$) inter-boundary tunneling events may take place and the only terms which preserve time-reversal symmetry\cite{zhang_ti_2006} can be distinguished in spin-preserving and spin-flipping tunneling described by the following contributions to the Hamiltonian:
\begin{eqnarray}
\label{eq:backscattering}
H_{sp}=\sum_{\sigma=\uparrow,\downarrow} \int dx &&\lbrack \Gamma_{sp}(x) \psi^\dagger_{R\sigma}(x)\psi_{L\sigma}(x)+\nonumber \\
&&+\Gamma_{sp}(x)^\ast \psi^\dagger_{L\bar{\sigma}}(x)\psi_{R\bar{\sigma}}(x)
\rbrack,
\end{eqnarray}
\begin{eqnarray}
\label{eq:spin-flip}
H_{sf}=\sum_{\alpha=L,R} \int dx &&\xi_\alpha \lbrack \Gamma_{sf}(x) \psi^\dagger_{\alpha\uparrow}(x)\psi_{\alpha\downarrow}(x)+\nonumber \\
&&+\Gamma_{sf}(x)^\ast \psi^\dagger_{\alpha\downarrow}(x)\psi_{\alpha\uparrow}(x)
\rbrack,
\end{eqnarray}
where $\alpha=\{L,R\}$,  $\xi_R=+1,\xi_L=-1$ is the chirality, while $\Gamma_{sf(sp)}(x)$ are the space-dependent tunneling amplitudes: $\Gamma_{sf(sp)}(x) = 2\hbar v_F\gamma_{sf(sp)}\delta(x)$.
Here the local modification of the spin-orbit coupling\cite{so-react} at $x=0$ governed by $\gamma_{sf}$ produces an effective spin-flipping mechanism.
Concerning the electron-electron interaction, the only terms that preserve the time-reversal symmetry of the system are the dispersive
\begin{eqnarray}
\label{eq:disp}
H_{d}=\sum_{\alpha=L,R} \int dx g_{2\bot}(x) :\rho_{\alpha\uparrow}(x)\rho_{\bar{\alpha}\downarrow}(x):,
\end{eqnarray}
and the forward scattering
\begin{eqnarray}
\label{eq:disp}
H_{f}=\sum_{\alpha=L,R; \sigma} \int dx \frac{g_{4\|}(x)}{2} :\rho_{\alpha\sigma}(x)\rho_{\alpha\sigma}(x):,
\end{eqnarray}
where $\rho_{\alpha \sigma}=:\psi^\dagger_{\alpha \sigma}\psi_{\alpha \sigma} :$ denotes the electron density with $\alpha=L,R$ and spin $\sigma$.
Thus, in Eq.~(\ref{eq:hamiltonian}), the Coulomb repulsion originated by unscreened charges at the QPC is accounted for $H_C=H_d+H_f$. The local nature of the interaction suggests that the coupling functions $g_{2\bot}(x)$, $g_{4\|}(x)$ are different from zero only at the QPC, while the case of pure Coulomb repulsion requires $g_{2\bot}(x)=g_{4\|}(x)>0$. Accordingly, we assume $g_{2\bot}(x)=g_d \mathcal{W}\delta(x)$ and $g_{4\|}(x)=g_f \mathcal{W}\delta(x)$, with $g_d=g_f\equiv g$, where $\mathcal{W}$ is the area of the QPC.

In the presence of side gates $V_{g}$ acting at top and bottom sides of the QPC, an additional term appears in the Hamiltonian:
\begin{eqnarray}
\label{eq:gates}
H_g=\sum_{\alpha,\sigma}\int dx e V_{g} (x)\rho_{\alpha \sigma},
\end{eqnarray}
where the local action of the gate is modeled by $V_g(x)=V_g\mathcal{W}^{\frac{1}{2}}\delta(x)$, being $\mathcal{W}^{\frac{1}{2}} \sim l$ the length of the QPC.\\
As a final comment, we notice that the delta-like nature of the potentials introduced in $H_C$ does not limit the generality of our model as long as the low energy regime is considered. Indeed, an incoming electron of momentum $k_E=E/(\hbar v_F)$ is not able to resolve the internal structure of the QPC region if the relation $k_E l/(2\pi) \ll1$ is fulfilled. Thus within the low energy regime ($E\ll hv_F/l\equiv E_T$) our results are not affected by the simplified assumptions we made, while beyond this regime model-dependent features are expected. However, the energy scale $E_T$, estimated by assuming $v_F \approx 3.2 \cdot 10^{5}$ m/s and $l=20$ nm, takes a value which is several times bigger than the topological insulator bulk gap $\Delta_{bulk}$ ($E_T \sim 1.3/l_{nm}$ eV, $l_{nm}$ being the QPC length expressed in nm). Thus the low energy condition which ensures the generality of the model is always satisfied as long as we consider the energy range in which the topological edge states represent the relevant modes involved in the charge transport.

\subsection{Symmetry-preserving mean-field approximation of $H_C$}
\label{ssec:spmfapp}
The model of corner junction described so far contains the term $H_C$ which is difficult to be treated within the scattering matrix formalism. Since we are interested in describing the Coulomb blockade physics, which is a fundamentally classical effect, it is sufficient a mean field treatment of the interaction. In performing this, however, a symmetry preserving approximation is required. Indeed, the original Hamiltonian describes a time reversal invariant system, which is a property to be preserved after the approximation has been made. In general the mean field treatment substitutes the interaction $H_C$ with a single particle operator of the form $H_{C}^{mf} \sim \lambda_1 \rho_{R\uparrow}+\lambda_2 \rho_{R\downarrow}+\lambda_3 \rho_{L\uparrow}+\lambda_4 \rho_{L\downarrow}$. The time reversal operator $\mathcal{T}=\tau_{x}\otimes i\sigma_{y} \mathcal{C}$, being
 $\mathcal{C}$ the conjugation operator, produces the field transformation $\mathcal{T}\psi_{\alpha\sigma}\mathcal{T}^{-1}=\zeta_{\sigma}\psi_{\bar{\alpha}\bar{\sigma}}$, with $\zeta_{\uparrow (\downarrow)}=+1(-1)$, and the notation $\bar{\alpha}$ ($\bar{\sigma}$) denoting the chirality (spin projection) opposite to $\alpha$ ($\sigma$). As a consequence the approximated interaction Hamiltonian preserves the time reversal invariance, i.e. $\mathcal{T}H_{C}^{mf}\mathcal{T}^{-1}=H_{C}^{mf}$, provided that $\lambda_1=\lambda_4$ and $\lambda_2=\lambda_3$. The standard mean field approximation of $H_C$ leads to the single particle operator $H_C^{mf} \approx \int dx \sum_{\alpha \sigma}\mathcal{U}_{\alpha \sigma}(x) \rho_{\alpha\sigma}(x)$, where \begin{equation}
\mathcal{U}_{\alpha \sigma}(x)=g_{2\bot}(x)\langle \rho_{\bar{\alpha} \bar{\sigma}}\rangle+g_{4\|}(x)\langle \rho_{\alpha \sigma}\rangle
\end{equation}
are electrostatic potentials describing the Coulomb repulsion originated by the electron densities $\langle \rho_{\alpha\sigma}\rangle$ at the QPC. These potentials, to be determined self-consistently, have the appropriate symmetry properties (i.e. $\mathcal{U}_{R\uparrow}(x)=\mathcal{U}_{L\downarrow}(x)$ and $\mathcal{U}_{R\downarrow}(x)=\mathcal{U}_{L\uparrow}(x)$) to make $H_C^{mf}$ a time reversal preserving approximation of $H_C$. Thus in the following we make the substitution $H_C\rightarrow H_{C}^{mf}$.

\section{Scattering fields approach}
\label{sec:scattering-field}
We now formulate a scattering field theory \textit{\`{a} la} B\"{u}ttiker\cite{buttiker_92} able to describe coherent spin and charge transport in the system shown in Fig.~\ref{fig:fig1} (a).\\
The charge or spin current operators $\hat{J}_{c/s}$ in first quantization are written as follows:
\begin{eqnarray}
\hat{J}_{c} &=& v_F e\hat{\tau}_z\otimes \mathbb{I}_{2 \times 2}\\\nonumber
\hat{J}_{s} &=& v_F \frac{\hbar}{2}\hat{\tau}_z\otimes \hat{\sigma}_{z},
\end{eqnarray}
where $\hat{\sigma}_{z}(\hat{\tau}_{z})$ stands for the Pauli matrix, $ \mathbb{I}_{2 \times 2}$ is the identity matrix acting on the spin sub-space of the Hilbert space given by the tensor product $| \alpha\rangle\otimes|\sigma\rangle$ ($\alpha \in \{R,L\}$, $\sigma \in \{\uparrow,\downarrow\}$).
To build a scattering field theory, one first defines the scattering field corresponding to each terminal $i=1,\ldots,4$ in terms of the incoming ($\hat{a}_{\alpha\sigma}(E)$) and outgoing ($\hat{b}_{\alpha\sigma}(E)$) electron operators, according to:
\begin{eqnarray}
\hat{\Psi}_1(x,t) &=& \int dE \frac{e^{-iEt/\hbar}}{\sqrt{h v_F}}\Bigl[\hat{a}_{R\downarrow}(E;x)+\hat{b}_{L\uparrow}(E;x)\Bigl]\\\nonumber
\hat{\Psi}_2(x,t) &=& \int dE \frac{e^{-iEt/\hbar}}{\sqrt{h v_F}}\Bigl[\hat{a}_{R\uparrow}(E;x)+\hat{b}_{L\downarrow}(E;x)\Bigl]\\\nonumber
\hat{\Psi}_3(x,t) &=& \int dE \frac{e^{-iEt/\hbar}}{\sqrt{h v_F}}\Bigl[\hat{a}_{L\downarrow}(E;x)+\hat{b}_{R\uparrow}(E;x)\Bigl]\\\nonumber
\hat{\Psi}_4(x,t) &=& \int dE \frac{e^{-iEt/\hbar}}{\sqrt{h v_F}}\Bigl[\hat{a}_{L\uparrow}(E;x)+\hat{b}_{R\downarrow}(E;x)\Bigl],
\end{eqnarray}
where $\hat{a}_{\alpha\sigma}(E;x)=\hat{a}_{\alpha\sigma}(E)| \alpha\rangle\otimes|\sigma\rangle \exp(i\eta_{\alpha} k_E x)$ with $\eta_{R}=-\eta_{L}=1$ and the wavevector $k_E=E/(\hbar v_F)$ (and similarly for $\hat{b}$) .\\
Then, the second-quantized current operators $\hat{\mathbf{J}}^{(i)}_{c/s}=\hat{\Psi}_i^{\dag}\hat{J}_{c/s}\hat{\Psi}_i$ in the terminal $i$ are explicitly given by ($\nu \in \{c,s\}$):
\begin{equation}
\label{eq:current_op}
\hat{\mathbf{J}}^{(i)}_{\nu}=\epsilon_i g_{\nu}\Bigl[ (\xi_{\nu})^{i+1}\hat{a}^{\dag}_{i}\hat{a}_{i}+(\xi_{\nu})^{i}\hat{b}^{\dag}_{i}\hat{b}_{i}\Bigl],
\end{equation}
where $g_{c}=|e|/h$, $g_{s}=1/(4\pi)$, $\xi_{c/s}=\mp 1$ and $\epsilon_{1,4}=-1=-\epsilon_{2,3}$. In writing Eq.~(\ref{eq:current_op}) we made use of the Fourier transform $\hat{a}_i(t)=\int dE \hat{a}_{i}(E)\exp[-iEt/\hbar]$, while the following correspondence has been made:
$[b_1, b_2, b_3, b_4]^t=[b_{L\uparrow}, b_{L\downarrow}, b_{R\uparrow}, b_{R\downarrow}]^t$, $[a_1, a_2, a_3, a_4]^t=[a_{R\downarrow}, a_{R\uparrow}, a_{L\downarrow}, a_{L\uparrow}]^t$.\\
The expectation value $\langle \hat{\mathbf{J}}^{(i)}_{\nu}\rangle$ can be computed making use of the scattering relation $b_j=\sum_{i}S_{ji}a_i$ and from quantum statical average $\langle \hat{a}^{\dag}_{j}(E) \hat{a}_{i}(E')\rangle=\delta_{ij}\delta(E-E')f_i(E)$, being $f_i(E)$ the Fermi-Dirac distribution with electrochemical potential $\mu_i=\mu+eV_i$. After direct computation we get:
\begin{equation}
\label{eq:current_exp}
\langle\hat{\mathbf{J}}^{(i)}_{\nu}\rangle=\epsilon_i g_{\nu}\int dE \sum_{j}\Bigl[\delta_{ij}(\xi_{\nu})^{i+1}+(\xi_{\nu})^i|S_{ij}(E)|^2\Bigl]f_{j}(E).
\end{equation}
The charge current fluctuations are described by the operators $\delta \hat{I}_i(t)=\hat{\mathbf{J}}^{(i)}_{c}(t)-\langle \hat{\mathbf{J}}^{(i)}_{c}(t)\rangle$ and can be computed within the scattering formalism. In this context $\langle\delta \hat{I}_i(t)\rangle=0$, while the current-current correlation function\cite{khlus87}
\begin{equation}
K_{ij}(\tau)=\frac{1}{2}\langle\{\delta \hat{I}_i(t),\delta \hat{I}_j(t+\tau)\}\rangle
\end{equation}
contains non-local information on the spectral density of the current fluctuations, $\mathcal{S}_{ij}(\Omega)=2K_{ij}(\Omega)$, being the Fourier transform defined as $K_{ij}(\Omega)=\int d\tau K_{ij}(\tau)\exp[i\Omega \tau]$. The non-local spectral density
\begin{eqnarray}
\label{eq:spectral_density}
&&\mathcal{S}_{ij}(\tau)=\int dE dE' \exp \Bigl[i\frac{E'-E}{\hbar}\tau \Bigl]\times\\\nonumber
&&\sum_{m,\mu}\mathcal{A}^{(i)}_{m \mu}(E,E')\mathcal{A}^{(j)}_{\mu m}(E,E')f_{m}(E)(1-f_{\mu}(E'))+\\\nonumber
&&+(i \rightarrow j, j \rightarrow i ; \tau \rightarrow -\tau),
\end{eqnarray}
is explicitly written in terms of scattering matrix elements according to the relation:
\begin{equation}
\mathcal{A}^{(i)}_{\mu m}(E,E')=(-)^{i+1}\epsilon_{i}g_{c}[\delta_{i\mu}\delta_{im}-S^{\ast}_{i\mu}(E)S_{im}(E')].
\end{equation}
Eq.~(\ref{eq:spectral_density}) contains information on the fluctuation properties of the charge currents flowing through the junction and contains the equilibrium noise contribution (thermal noise) and the non-equilibrium (shot-noise) term. Thus effects of finite bias, temperature and frequency $\Omega$ are simultaneously taken into account in our formulation. Here it is worth to mention that in our notation the Kirchhoff's law is written as follows:
\begin{equation}
\label{eq:KL}
\langle\hat{\mathbf{J}}^{(1)}_{c}\rangle+\langle\hat{\mathbf{J}}^{(2)}_{c}\rangle-\langle\hat{\mathbf{J}}^{(3)}_{c}\rangle-\langle\hat{\mathbf{J}}^{(4)}_{c}\rangle=0,
\end{equation}
due to our choice of the area vectors encircling the lead regions. Eq.~(\ref{eq:KL}) takes the form $\sum_{i}\langle\hat{\mathbf{J}}^{(i)}_{c}\rangle=0$ when the substitution $\hat{\mathbf{J}}^{(i)}_{c}\rightarrow \epsilon_{i}(-)^{i}\hat{\mathbf{J}}^{(i)}_{c}$ is made. As the effect of the mentioned substitution the spectral density is redefined accordingly: $\mathcal{S}_{ij}(\tau)\rightarrow \epsilon_{i}\epsilon_{j}(-)^{i+j}\mathcal{S}_{ij}(\tau)$. Although the physical equivalence of the area vector conventions, in the following we will work within the new conventions ($[\hat{\mathbf{J}}^{(i)}_{c}]_{new}= \epsilon_{i}(-)^{i}[\hat{\mathbf{J}}^{(i)}_{c}]_{old}$,  $[\mathcal{S}_{ij}(\tau)]_{new}=\epsilon_{i}\epsilon_{j}(-)^{i+j}[\mathcal{S}_{ij}(\tau)]_{old}$) to better compare our results with the existing literature.\\
In order to capture non-equilibrium charging effects at the QPC a self-consistent evaluation of $\mathcal{U}_{\alpha \sigma}(x)$ is required. In our model, however, interaction effects are approximated by ultralocal potentials and thus the mean field terms $\mathcal{U}_{\alpha \sigma}(x)$ are determined by the electron densities $\langle \hat{\Psi}^{\dagger}_{i}(x,t) \hat{\Psi}_{i}(x,t)\rangle_{x=0}$ at the QPC. Within the scattering formalism the particle density at the QPC can be evaluated according to the formula:
\begin{equation}
\label{eq:nedistr}
\langle \hat{\Psi}^{\dagger}_{j}(x,t) \hat{\Psi}_{j}(x,t)\rangle_{x=0}=\int \frac{dE}{h v_F}f^{(j)}_{ne}(E),
\end{equation}
where the non-equilibrium distribution $f^{(j)}_{ne}(E)=\sum_{i}[\delta_{ji}+|S_{ji}(E)|^2]f_{i}(E)$ accounts for the perturbation of the occupation number of a scattering state due to the presence of the QPC, while $(h v_F)^{-1}$ represents the one dimensional density of states. As will be clear in the following discussion, the scattering matrix of the problem is energy-independent and all the integrals present in the theory can be performed analytically. In the following discussion, however, simplifying approximations are made in order to put in evidence the physics behind the formalism. In particular, in the experimental relevant temperature range (low temperature regime) Eq.~(\ref{eq:nedistr}) is well approximated by its zero temperature value\cite{nota_int_fermi} and thus the charge density imbalance at the QPC takes the form:
\begin{equation}
\label{eq:charge_imb}
\delta \mathcal{Q}^{(i)}=C_{Q}[V_{i}+\sum_{j}|S_{ij}|^2V_{j}],
\end{equation}
where  $C_{Q}=e^2/(h v_F)$ is the quantum capacitance of the junction, while $\delta \mathcal{Q}^{(i)}=-|e|(\langle \hat{\Psi}^{\dagger}_{i}(x,t) \hat{\Psi}_{i}(x,t)\rangle_{x=0}-\rho_0)$ represents the charge density imbalance induced by a shift of the electron density with respect to the equilibrium value $\rho_0=2(\mu+\Delta_{bulk}/2)/(hv_F)$, where $\mu$ is the chemical potential and $\Delta_{bulk}$ is the bulk gap. Eq.~(\ref{eq:charge_imb}) shows that changing the electrochemical potential of the four terminals of the same amount $eV_{j}\equiv eV$ produces the charge density imbalance $\delta \mathcal{Q}=2C_Q V$. This is equivalent to the field effect produced by a back-gate acting below the whole hetero-structure.

\subsection{Boundary conditions and self-consistent computation of the scattering matrix}
\label{ssec:bcs-self}
The scattering matrix $S_{ij}$ is a four by four unitary matrix whose diagonal entries vanish by helicity and time-reversal symmetry while all the other entries
can be explicitly determined as a function of the tunneling amplitudes $\gamma_{sp},\gamma_{sf}$ by imposing the proper boundary conditions (BCs) on the wave functions\cite{Dolcini11,Citro11,Romeo12}. Since the wave functions are continuous in the regions $x<0$ and $x>0$, we only have to impose the matching conditions where Dirac delta potentials are present, i.e. at $x=0$. By using the equation of motion of the quantum field $\psi_{\alpha\sigma}(x)$, i.e.
\begin{eqnarray}
&&\Bigl[-i\hbar v_F \xi_{\alpha}\partial_{x}+2\hbar v_F \delta(x)~u_{\alpha\sigma}(x)-E\Bigl]\psi_{\alpha\sigma}(x)+\\\nonumber
&&+2\hbar v_F\Bigl[\gamma_{sp}\psi_{\bar{\alpha}\sigma}(x)+\gamma_{sf}\xi_{\alpha}\psi_{\alpha\bar{\sigma}}(x)\Bigl]\delta(x)=0,
\end{eqnarray}
with
\begin{equation}
u_{\alpha \sigma}(x)=u\Bigl[ \langle\rho_{\bar{\alpha}\bar{\sigma}}(x)\rangle+\langle\rho_{\alpha\sigma}(x)\rangle\Bigl]+u_{g},
\end{equation}
being $u=\frac{g\mathcal{W}}{2\hbar v_F}$ the interaction parameter [notice that $g_{2\perp}(x)=g_{4 \parallel}(x)=g \mathcal{W}\delta(x)$] and $u_{g}=\frac{e V_{g}\sqrt{\mathcal{W}}}{2\hbar v_F}$, and explicitly taking into account the properties of the Dirac delta potential under integration, one obtains the following  matching conditions ($\psi_{\alpha\sigma}(x=0^{\pm})\equiv \psi_{\alpha\sigma}^{(\pm)}$, $u_{\alpha\sigma}(x=0^{\pm})\equiv u_{\alpha\sigma}^{(\pm)}$):
\begin{eqnarray}
\label{eq:BCs}
&-&i\xi_{\alpha}[\psi^{(+)}_{\alpha\sigma}-\psi^{(-)}_{\alpha\sigma}]+\gamma_{sp}[\psi^{(+)}_{\bar{\alpha}\sigma}+\psi^{(-)}_{\bar{\alpha}\sigma}]+\\\nonumber
&+&\xi_{\alpha}\gamma_{sf}[\psi^{(+)}_{\alpha\bar{\sigma}}+\psi^{(-)}_{\alpha\bar{\sigma}}]+u^{(+)}_{\alpha\sigma}\psi^{(+)}_{\alpha\sigma}+u^{(-)}_{\alpha\sigma}\psi^{(-)}_{\alpha\sigma}=0.
\end{eqnarray}
In Eq.~(\ref{eq:BCs}) we make the substitution $u_{\alpha \sigma}^{(\pm)}\rightarrow u_{\alpha\sigma}= [u_{\alpha \sigma}^{(+)}+u_{\alpha \sigma}^{(-)}]/2$ which is equivalent to consider the spatial average $u_{\alpha\sigma}$ of the self-generated Coulomb potential computed by using the values taken just after and before the constriction. Due to the time reversal symmetry we have $u_{R\uparrow}=u_{L\downarrow}=u_{1}$ and $u_{R\downarrow}=u_{L\uparrow}=u_{2}$, while the mean field self-consistency is provided by the equations
\begin{eqnarray}
\label{eq:self}
&& u_{1}=\frac{u}{2}\sum_{\mu=\pm}[\langle\rho_{R\uparrow}\rangle^{(\mu)}+\langle\rho_{L\downarrow}\rangle^{(\mu)}]+u_g\\\nonumber
&& u_{2}=\frac{u}{2}\sum_{\mu=\pm}[\langle\rho_{R\downarrow}\rangle^{(\mu)}+\langle\rho_{L\uparrow}\rangle^{(\mu)}]+u_g.
\end{eqnarray}
The explicit form of Eqs.~(\ref{eq:self}) in terms of the scattering matrix elements is given by the nonlinear equations in the unknown parameters $u_{1}$ and $u_{2}$:
\begin{equation}
\label{eq:self2}
\left(
  \begin{array}{c}
    u_{1} \\
    u_{2} \\
  \end{array}
\right)=(u\rho_0+u_g)\left(
                   \begin{array}{c}
                     1 \\
                     1 \\
                   \end{array}
                 \right)+\mathcal{M}\left(
                                      \begin{array}{c}
                                        v_{1} \\
                                        v_{2}  \\
                                        v_{3}  \\
                                        v_{4}  \\
                                      \end{array}
                                    \right),
\end{equation}
being the $\mathcal{M}$ matrix expression reported in Appendix \ref{app:A}.
Eqs.~(\ref{eq:self2}) must be solved self-consistently since the matrix elements of $\mathcal{M}$ depends on $u_1$ and $u_{2}$ themselves through the scattering matrix elements (see Eq.~(\ref{eq:m}) of Appendix \ref{app:A}). The dimensionless parameters $v_{i}$ represent the electrochemical potential shift imposed to the i-th lead as the effect of the voltage bias and are defined by $v_{i}=|e|V_{i}/(\mu+\frac{\Delta_{bulk}}{2})$, while the dimensionless quantity $u\rho_0+u_g$ is related to the field effect induced by the side-gate $V_{g}$. The scattering matrix elements, which completely define the self-consistency relations, can be determined making the substitutions
\begin{equation}
\left(
  \begin{array}{c}
    \psi_{R\uparrow}^{(+)} \\
   \psi_{R\downarrow}^{(+)} \\
    \psi_{L\uparrow}^{(+)} \\
    \psi_{L\downarrow}^{(+)} \\
  \end{array}
\right)\rightarrow\left(
                    \begin{array}{c}
                      b_{3} \\
                       b_{4} \\
                       a_{4} \\
                       a_{3} \\
                    \end{array}
                  \right),~~~\left(
  \begin{array}{c}
    \psi_{R\uparrow}^{(-)} \\
   \psi_{R\downarrow}^{(-)} \\
    \psi_{L\uparrow}^{(-)} \\
    \psi_{L\downarrow}^{(-)} \\
  \end{array}
\right)\rightarrow\left(
                    \begin{array}{c}
                      a_{2} \\
                       a_{1} \\
                       b_{1} \\
                       b_{2} \\
                    \end{array}
                  \right)
                 \end{equation}
within the Eqs.~(\ref{eq:BCs}) also taking into account the scattering relation $b_{i}=\sum_{j}S_{ij}a_{j}$. The scattering matrix $S[u_{1},u_{2},\gamma_{sp},\gamma_{sf}]$ determined according to the above procedure is given in Eq.~(\ref{eq:sm}) of Appendix \ref{app:B} and parametrically depends on the self-consistency parameters $u_{1}$ and $u_{2}$ to be determined by solving the nonlinear problem given by Eqs.~(\ref{eq:self2}). The scattering matrix $S[u_{1},u_{2},\gamma_{sp},\gamma_{sf}]$ depends explicitly on the details of the QPC encoded by the tunnel probabilities $\gamma_{sf/sp}$ and presents an hidden dependence on the voltage bias. Importantly, it does not depend on the energy of the scattering processes. It is also worth to mention that Eq.~(\ref{eq:sm}) provides a system description beyond the linear response theory and correctly reproduces the limit of negligible interaction (i.e. $u_{1/2} \rightarrow 0$)\cite{Dolcini11} and  of uncoupled boundaries (i.e. $\gamma_{sf}\rightarrow 0$, $\gamma_{sp}\rightarrow 0$). In the latter case the scattering matrix describes  the reflectionless transport of helical states moving along separate boundaries and its non-vanishing elements are simply the phase factors $S_{14}=S_{41}=-\exp(2i \arctan(1/u_{2}))$ and $S_{32}=S_{23}=-\exp(2i \arctan(1/u_{1}))$.
In particular the renormalization of the tunneling amplitudes promoted by the interaction is well captured by the present formalism. For instance, supposing $\gamma_{sf}=0$ and assuming $\gamma_{sp}$ as a small perturbation, we get $|S_{14}|^2=|S_{23}|^2\approx 1-4\tilde{\gamma}_{sp}^{2}+o(\gamma_{sp}^{4})$ and $|S_{12}|^2=|S_{34}|^2\approx 4\tilde{\gamma}_{sp}^{2}+o(\gamma_{sp}^{4})$ where the renormalized tunneling parameter is related to its bare value by the relation $\tilde{\gamma}_{sp}=\gamma_{sp}/\sqrt{(1+u_{1}^2)(1+u_{2}^2)}$. The renormalized tunneling  describes a lowering of inter-edges scattering probability induced by the interaction (i.e. $\tilde{\gamma}_{sp}<\gamma_{sp}$).\\
This behavior is in qualitative agreement with a renormalization group argument on the spin preserving and spin-flipping scattering within a helical Luttinger liquid model\cite{ferraro13,giamarchi-book}. When these terms become relevant, the tunneling amplitudes depend as a power-law on the Luttinger parameter, which is linear in the Coulomb interaction $U$, for small $U$.\\
Once the scattering matrix has been determined, the charge current and its fluctuation can be computed.

\section{Results}
\label{sec:results}
In the following we present the current-voltage characteristics and the current fluctuations assuming that the chemical potential is located in the middle of the topological insulator bulk gap, i.e. $\mu=0$. As a consequence, the energy shift imposed by the voltage bias to the local electrochemical potentials takes the form $|e|V_{j}=\frac{\Delta_{bulk}}{2}v_{j}$, while the dimensionless energy shifts $v_j$ are chosen in the interval $]-1,1[$, the latter condition ensuring that only topological edge states are involved in the charge transport. Within a real topological heterostructure going beyond this limit implies that the charge current is partially supported by states (not included in the theory) having energy above the insulating gap.\\
For a QPC of nanometric size ($\sqrt{\mathcal{W}}\sim 20$ nm), the dimensionless parameter $\rho_{0}u=\pi g \mathcal{W}\Delta_{bulk}/(hv_{F})^{2}$ can be estimated to be of order $\sim 0.5-5$. The latter estimate comes from identifying $g$ with the charging energy $e^2/(2C)$, being $C$ the capacitance of the nanometric constriction varying in the aF range.

\subsection{Current-voltage characteristics and differential conductance}
\label{ssec:iv-diff-cond}
We study the current flowing through the system assuming the $T=0$ limit which is appropriate to describe experiments performed in thermal bath with temperature ranging from tens of mK to few Kelvin. Under this condition, since the scattering matrix does not depend on the energy, the charge current $I_{ch}^{(i)}$ flowing through the i-th lead can be written as
\begin{equation}
\label{eq:charge_curr}
I_{ch}^{(i)}=I_{0}\sum_{j}\Bigl[\delta_{ij}-|S_{ij}|^2\Bigl]v_{j},
\end{equation}
where $I_{0}=|e|\Delta_{bulk}/(2h)$ is used as current unit. In particular, if the value of the bulk gap is $\delta~\mathrm{meV}$, $I_{0}$ takes values in the nA range, i.e. $\sim 19.3~\mathrm{nA}\times~\delta$. Despite the linear response form of Eq.~(\ref{eq:charge_curr}), the hidden dependence on the applied voltages of the scattering matrix determines non-linear current-voltage characteristics. The deviation from the ohmic behavior and the current suppression at low bias  originates from the Coulomb repulsion at the QPC and represents a clear signature of the Coulomb blockade regime. Since the electron density within the constriction can be tuned by using the side-gate $u_g$, the Coulomb blockade regime can be switched on and off by using all-electrical means\cite{CB_TIexp2012}. In order to put in evidence the renormalization effects induced by the Coulomb interaction, we focus on a crossed voltage-bias configuration for which the leads $1$ and $3$ are grounded (i.e. $V_{1}=V_{3}=0$), while a finite voltage drop $V$ is applied to the leads $2$ and $4$, i.e. $V_{2}=V/2$ and  $V_{4}=-V/2$. Accordingly, the explicit expression of the charge currents $I_{ch}^{(1)}$ and $I_{ch}^{(2)}$ in terms of the scattering matrix elements can be written in the form:
\begin{eqnarray}
\label{eq:curr-scatt1}
&&I_{ch}^{(1)}=\frac{e^2}{h}\Bigl[|S_{14}|^2-|S_{12}|^2\Bigl]\frac{V}{2}\\
\label{eq:curr-scatt2}
&&I_{ch}^{(2)}=\frac{e^2}{h}\Bigl[1+|S_{24}|^2\Bigl]\frac{V}{2},
\end{eqnarray}
while the charge conservation implies that $I_{ch}^{(3)}=-I_{ch}^{(1)}$ and $I_{ch}^{(2)}=-I_{ch}^{(4)}$. The usefulness of the crossed bias configuration in recognizing the renormalized junction parameters is evidenced
in Eqs.~(\ref{eq:curr-scatt1})-(\ref{eq:curr-scatt2}). Indeed, one expects that $I_{ch}^{(2)}$ is essentially linear with respect to the applied
bias since $|S_{24}|^2\sim \gamma_{sf}^2$ is only a weak perturbation, while important renormalization effects and nonlinear behavior are
expected for $I_{ch}^{(1)}$.
\begin{figure}[!h]
\includegraphics[scale=0.6]{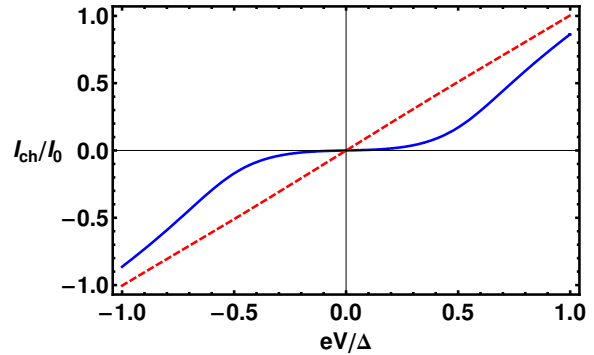}
\caption{(color online) $I_{ch}^{(1)}$ (full line) and $I_{ch}^{(2)}$ (dashed line) as a function of the voltage drop $V$. The remaining model parameters have been fixed as follows: $\gamma_{sp}=0.4$, $\gamma_{sf}=0.1$, $\rho_0 u=3$ and $u_{g}=-3$. The charge conservation implies $I_{ch}^{(3)}=-I_{ch}^{(1)}$ and $I_{ch}^{(2)}=-I_{ch}^{(4)}$, so that $\sum_{j}I_{ch}^{(j)}=0$ is respected. Here and in Figs.~3,~4,~7,~8 we set $\Delta \equiv \Delta_{bulk}$ to shorten the axis label.}
\label{fig:fig2}
\end{figure}
In fact, in the limit of small inter-edge coupling, i.e. $\gamma_{sp}\ll 1$,  and $\gamma_{sf}=0$,  one obtains
\begin{eqnarray}
&&I_{ch}^{(1)}\approx\frac{e^2}{h}\Bigl[1-8\tilde{\gamma}_{sp}^2\Bigl]\frac{V}{2}\\
&&I_{ch}^{(2)}\approx\frac{e^2}{h}\frac{V}{2}.
\end{eqnarray}
The functional form of $I_{ch}^{(1)}$ is useful in identifying the renormalized tunneling amplitude $\tilde{\gamma}_{sp}^2$ as a function of the applied bias when experimental data are analyzed. This
 is confirmed by Fig.~\ref{fig:fig2}, where the charge currents $I_{ch}^{(1)}$ (full line) and $I_{ch}^{(2)}$ (dashed line) are represented as a function of $V$, while fixing the remaining parameters as shown in the figure caption. The $I_{ch}^{(1)}$ \textit{vs} $V$ curve shows a clear Coulomb blockade region at low bias, while an ohmic behavior is detected for $I_{ch}^{(2)}$.
\begin{figure}[!h]
\includegraphics[scale=0.5]{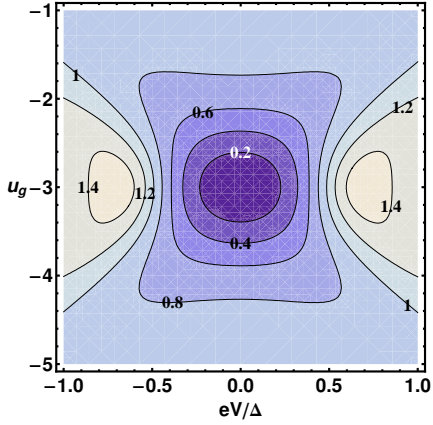}
\caption{(color online) Contour plot of the dimensionless differential conductance $G=(e^2/h)^{-1}\frac{d I_{ch}^{(1)}}{d(V/2)}$ as a function of the applied voltage $eV/\Delta$ and of the side-gate voltage $u_g$. The junction parameters are: $\gamma_{sp}=0.4$, $\gamma_{sf}=0.1$, $\rho_{0}u=3$. Darker areas represent lower conductance values, while each contour line is labeled by the corresponding conductance value. At low bias, the Coulomb Blockade region is evident.}
\label{fig:fig3}
\end{figure}
To further characterize the Coulomb blockade regime, in Fig.~\ref{fig:fig3} we show the normalized differential conductance $G=(e^2/h)^{-1}\frac{d I_{ch}^{(1)}}{d(V/2)}$ as a function of the applied voltage $eV/\Delta$ and of the side-gate voltage $u_g$, while the remaining parameters are fixed as done in Fig.~\ref{fig:fig2}. The minimum conductance value is
obtained for $V=0$ and $u_{g}=-3$, while the conductance progressively increases and the linearity of the current-voltage curve
is gradually recovered as the voltage of the side gate $u_g$ is moved from the value -3. We have also verified that for different values of $\rho_0 u$ the Coulomb blockade region is always centered at a side-gate value $u_{g}=-\rho_0 u$. This feature is evident in Fig.~\ref{fig:fig4} where the differential conductance $G$ is shown as a function of the bias voltages for $\rho_{0} u=0.5$ (upper left panel), $\rho_{0} u=1$ (upper right panel), $\rho_{0} u=2$ (lower left panel) and $\rho_{0} u=3$ (lower right panel), while the junction parameters have been fixed as $\gamma_{sp}=0.4$, $\gamma_{sf}=0.1$. At low bias ($V \approx 0$), the Coulomb blockade regime is induced by side-gate voltages $u_{g} \in [-3\rho_{0}u/2, -\rho_{0}u/2]$, and the extension of the Coulomb blockade region is of the order
of $\rho_{0}u$ along the $u_g$ axis. The Coulomb blockade regime vanishes for side-gate values fulfilling the condition $|u_{g}+\rho_{0}u|>\rho_{0}u/2$. Apart from specific aspects related to the multi-terminal nature of the device, here we notice that $\rho_{0}u/2 \propto g \sim e/C$ plays
the same role of the critical voltage $V_c=e/C$ defining the Coulomb diamond extension in the conventional Coulomb blockade theory. Furthermore, despite the zero-temperature limit considered here, the transition from Coulomb blockade and conducting regime as a function of $V$ is not sharp (See for instance the behavior of $I_{ch}^{(1)}$ close to $eV/\Delta=0.5$ in Fig.~\ref{fig:fig2}). This behavior originates from the multi-terminal nature of the system which allows a non-thermal smearing of the current-voltage curves due to the current leakage towards a different electrode. Thus, the current leakage plays the same role of the thermal fluctuations (at finite temperature) in removing the Coulomb blockade.
\begin{figure}[!h]
\includegraphics[scale=0.5]{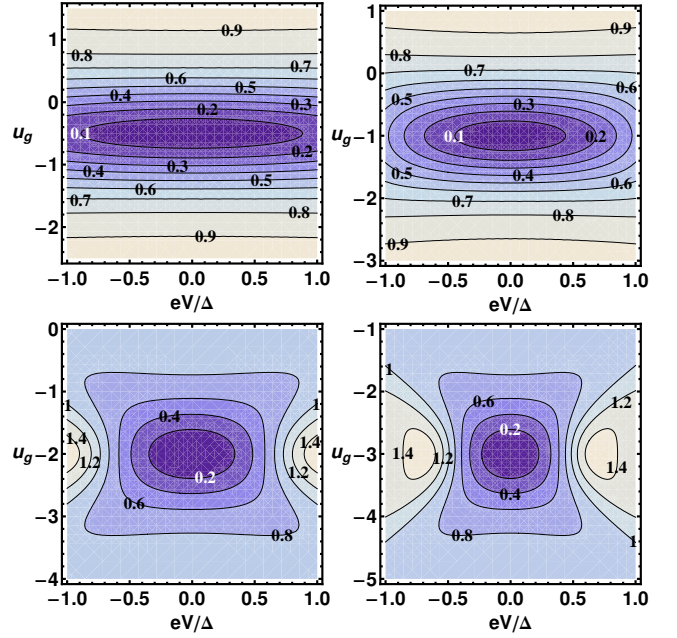}
\caption{(color online) Contour plot of the dimensionless differential conductance $G=(e^2/h)^{-1}\frac{d I_{ch}^{(1)}}{d(V/2)}$ as a function of the applied voltage $eV/\Delta$ and of the side-gate voltage $u_g$. The junction parameters are: $\gamma_{sp}=0.4$, $\gamma_{sf}=0.1$, while $\rho_{0} u=0.5$ for the upper left panel, $\rho_{0} u=1$ for the upper right panel, $\rho_{0} u=2$ for the lower left panel and $\rho_{0} u=3$ for the lower right panel. The Coulomb blockade regime is removed when $u_{g}$ verifies the condition $|u_{g}+\rho_{0}u|>\rho_{0}u/2$.}
\label{fig:fig4}
\end{figure}

\subsection{Current-Current correlations and noise}
\label{ssec:cur-cur-noise}
Exploiting the energy independence of the scattering matrix of the problem, we can write the finite frequency spectral density of the current fluctuation $\mathcal{S}_{ij}(\Omega)$ in the following form:
\begin{eqnarray}
\label{eq:spectral}
\mathcal{S}_{ij}(\Omega)&=&\frac{e^2}{h}\Bigl [ (\delta_{ij}-|S_{ji}|^2) \mathcal{F}_{ii}(\Omega)-|S_{ij}|^2 \mathcal{F}_{jj}(\Omega)\nonumber\\
&+& \sum_{mn}S^{\ast}_{im}S_{jm}S^{\ast}_{jn}S_{in}\mathcal{F}_{mn}(\Omega)\Bigl],
\end{eqnarray}
where we introduced the integral function
\begin{eqnarray}
\mathcal{F}_{mn}(\Omega)&=&\int_{-\Delta_{bulk}/2}^{\Delta_{bulk}/2} dE \Bigl[f_{m}(E)(1-f_{n}(E-\hbar \Omega))\nonumber\\
&+&f_{n}(E)(1-f_{m}(E+\hbar \Omega))\Bigl],
\end{eqnarray}
which can be analytically determined. The frequency resolved quantity $\mathcal{S}_{ij}(\Omega)$ [see Eq.~(\ref{eq:spectral})] takes simultaneously into account finite temperature and finite bias effects and presents a non-linear dependence with respect to the applied voltage. Thus, within this framework, non-equilibrium effects under non-linear response of the system can be fully accounted for. In the following we focus on the zero-frequency noise component $\mathcal{S}_{ij}(\Omega \rightarrow 0)$. Experimentally, $\mathcal{S}_{ij}(\Omega \rightarrow 0)$ is determined from the flat region of the spectrum (frequencies above few tens of kHz) where one can neglect the contribution of $1/f^{\gamma}$ noise, which is instead dominant at low frequency. Since the flicker noise completely masks any frequency dependence of $\mathcal{S}_{ij}(\Omega)$, the only experimentally accessible information is encoded in the zero frequency fluctuation.

\subsection{Thermal noise}
\label{ssec:thermal-noise}
In the absence of applied bias ($V_j=0$) thermally activated particles can gain a sufficient amount of energy to escape from a given electrode. As a consequence, within a given time interval, a particles number fluctuation is possible even in the absence of a net dc current. These equilibrium processes, which can be of local and non-local nature, are mathematically characterized by taking the zero-frequency limit of Eq.~(\ref{eq:spectral}) assuming a grounded configuration of the electrodes\cite{note2} ($V_j=0$). Using the time reversal symmetry ($|S_{ij}|^2=|S_{ji}|^2$) and the conservation of the helicity ($S_{ii}=0$), the local component of thermal noise detected in the ith lead can be written as
\begin{equation}
S_{th}\equiv\mathcal{S}_{ii}(\Omega=0)=S_{th}^{U}\frac{\exp(\frac{\Delta_{bulk}}{2 k_{B}T})-1}{\exp(\frac{\Delta_{bulk}}{2 k_{B}T})+1},
\end{equation}
while the non-local component ($i \neq j$) of the thermal fluctuation takes the form
\begin{equation}
\mathcal{S}^{(th)}_{ij}(\Omega=0)=-S_{th}^{U}|S_{ij}|^2\frac{\exp(\frac{\Delta_{bulk}}{2 k_{B}T})-1}{\exp(\frac{\Delta_{bulk}}{2 k_{B}T})+1}.
\end{equation}
Both the local and non-local components of the thermal fluctuation have a multiplicative universal factor $S_{th}^{U}=4k_BT(e^2/h)$ and
a factor depending on the insulating gap $\Delta_{bulk}$. At low temperature, however, $\frac{\exp(\frac{\Delta_{bulk}}{2 k_{B}T})-1}{\exp(\frac{\Delta_{bulk}}{2 k_{B}T})+1}$ approaches $1$ and thus the local component of the thermal noise becomes a universal quantity which does not depend on the conductance of the corner junction. The latter property is specific of the helical nature,
while in conventional materials (not-helical) the thermal noise is affected by the not universal value of the junction conductance. The temperature $T^{\ast}$ below which $S_{th}$ becomes universal depends on the value of the insulating bulk gap of the topological insulator. In Fig.~\ref{fig:fig4} we report the temperature
behavior of the local component of the thermal noise normalized to $S_{th}^{U}$ for $\Delta_{bulk}=4$ meV, $10$ meV, $30$ meV. The considered values of $\Delta_{bulk}$ are, respectively, appropriate for the two-dimensional topological insulators AlSb/InAs/GaSb/AlSb, CdTe/HgTe/CdTe and $\mathrm{Bi_2Se_3}$/graphene/$\mathrm{Bi_2Se_3}$ \cite{ando13,kuo13}.
\begin{figure}[!h]
\includegraphics[clip,scale=0.5]{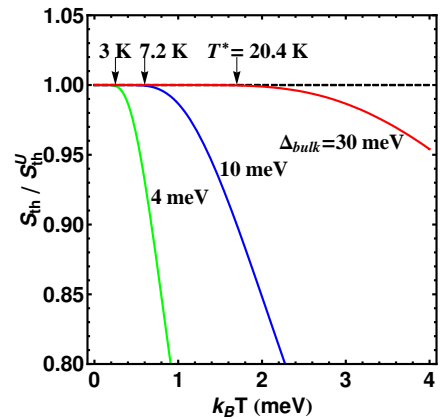}
\caption{(color online) Thermal noise of the ith lead compared to the universal value $S_{th}^{U}=4k_BT(e^2/h)$. The universal behavior of the thermal noise is reached below a certain temperature $T^{\ast}$ which is related to the value of the bulk insulating gap of the topological insulator assumed as energy cut-off of the one-dimensional theory. Different values of $\Delta_{bulk}$ are considered: $\Delta_{bulk}= 4$ meV, appropriate for AlSb/InAs/GaSb/AlSb;  $\Delta_{bulk}= 10$ meV, appropriate for CdTe/HgTe/CdTe; $\Delta_{bulk}= 30$ meV, appropriate for $\mathrm{Bi_2Se_3}$/graphene/$\mathrm{Bi_2Se_3}$\cite{ando13,kuo13}.}
\label{fig:fig5}
\end{figure}
As shown in Fig.~\ref{fig:fig5}, the universal character of $S_{th}$ becomes evident as the ratio $\Delta_{bulk}/(2k_{B}T)$ increases, i.e. when the bath temperature decreases below a certain gap-dependent temperature $T^{\ast}$ [\onlinecite{universalityth}]. In the case of a CdTe/HgTe/CdTe nanostructure ($\Delta_{bulk}=10$ meV), the helical nature of the system can be probed  at a bath temperature $T^{\ast}\approx 7.2$~K, which is much higher than the mK range usually required to verify the emergence of a topological phase in the quantum well. Accordingly, the measure of the thermal noise $S_{th}$ of the junction contains information on the helical nature of the system, the latter being already evident at bath temperature of few Kelvin.\\
Information on the junction parameters $\gamma_{sf/sp}$ can instead be deduced from the non-local component of the thermal fluctuations described by $\mathcal{S}^{(th)}_{ij}(\Omega=0)$, $i \neq j$. For instance, a corner junction with negligible reactivation of the spin-orbit coupling within the QPC region is described by $\gamma_{sf}=0$ and presents negligible non-local fluctuations between a couple of electrodes only connected by spin-flipping tunneling processes, i.e. $\mathcal{S}^{(th)}_{13}=\mathcal{S}^{(th)}_{31}=\mathcal{S}^{(th)}_{24}=\mathcal{S}^{(th)}_{42}=0$. On the other hand, for $\gamma_{sf} \neq 0$, the ratio between the correlation functions $\mathcal{S}^{(th)}_{13}/\mathcal{S}^{(th)}_{12}=(\gamma_{sf}/\gamma_{sp})^2$ is a temperature
independent quantity depending on the relative frequency of spin-preserving and spin-flipping tunneling events. At sufficiently low temperature $\mathcal{S}^{(th)}_{ij}/S_{th}^{U}=-|S_{ij}|^2$ and thus the non-local thermal fluctuation can be studied as a function of the side-gate voltage $u_{g}$. The latter analysis is performed in Fig.~\ref{fig:fig6} where $\mathcal{S}^{(th)}_{ij}/S_{th}^{U}=-|S_{ij}|^2$ \textit{vs} $u_g$ curves are shown.
\begin{figure}[!h]
\includegraphics[scale=0.6]{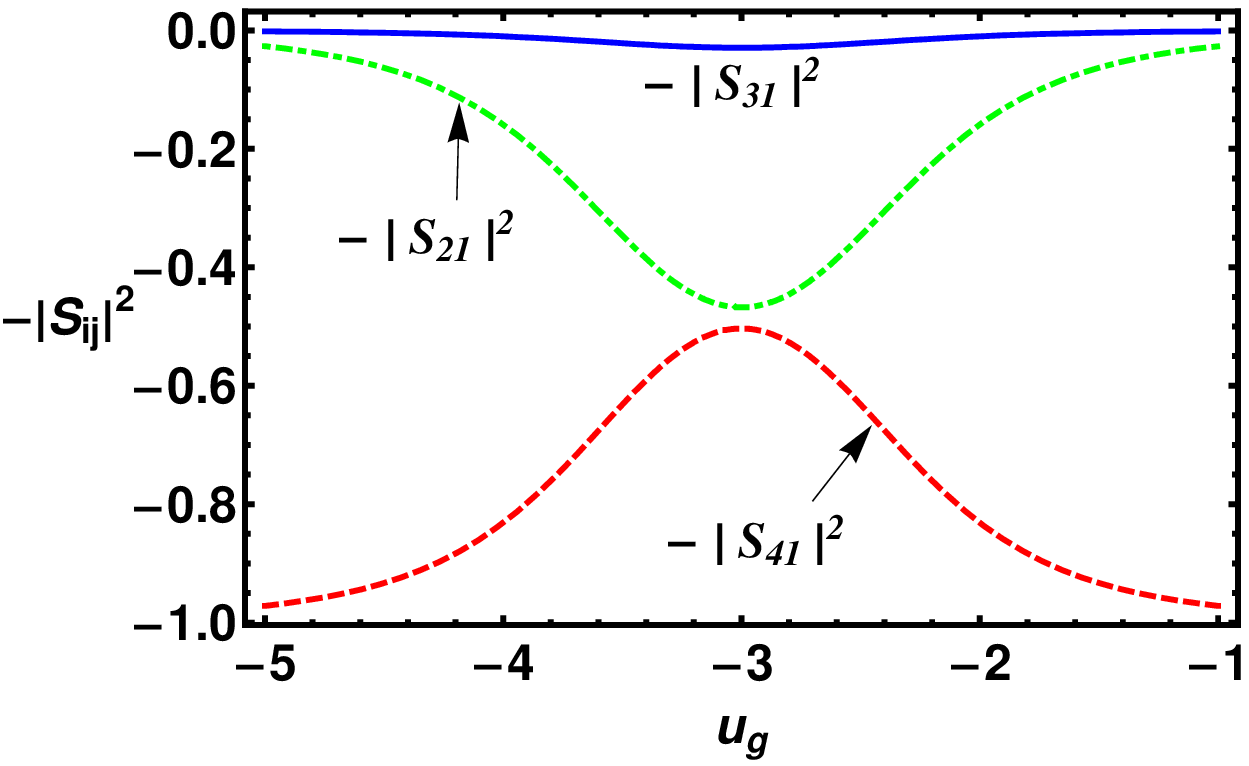}\\
\includegraphics[scale=0.6]{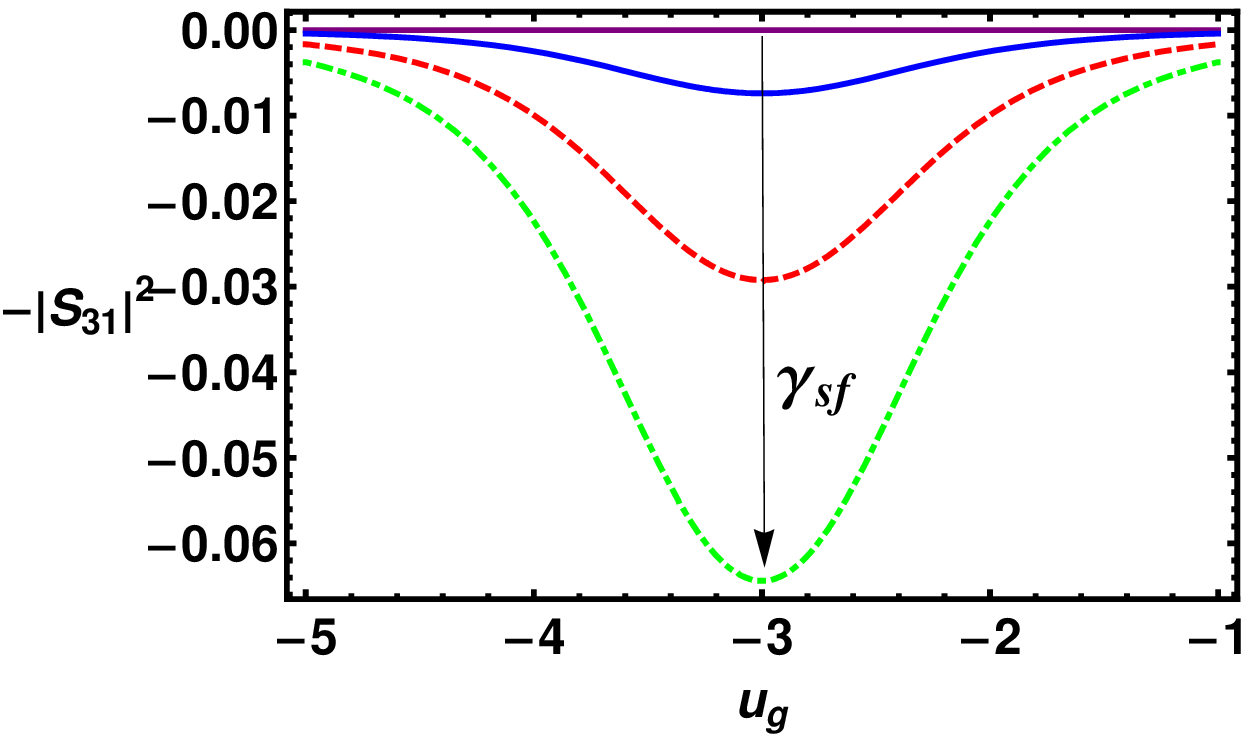}
\caption{(color online) Upper panel: $-|S_{21}|^2$, $-|S_{31}|^2$, $-|S_{41}|^2$ as a function of the side-gate voltage $u_g$. The junction parameters are: $\gamma_{sp}=0.4$, $\gamma_{sf}=0.1$, $\rho_{0}u=3$. Lower panel: $-|S_{31}|^2$ as a function of the side-gate voltage $u_g$. The different curves are obtained by fixing $\gamma_{sp}=0.4$ and $\rho_{0}u=3$, while setting the spin-flipping strength to $\gamma_{sf}=0$ (top curve), 0.05, 0.1, 0.15 (bottom curve).}
\label{fig:fig6}
\end{figure}
As a general remark, we observe that all the non-local correlations shown in Fig.~\ref{fig:fig6} are negative. This is consistent with the fact that in thermal equilibrium and at vanishing frequency $\Omega \rightarrow 0$, not only the particles flux is conserved but also its fluctuation. Since an increasing flux at one terminal must be compensated by a decreasing of the current at another terminal, non-local correlations must be negative (or at best zero). This argument fails under special conditions only in case of a bosonic system.\\
In the upper panel of Fig.~\ref{fig:fig6} we show $-|S_{21}|^2$, $-|S_{31}|^2$, $-|S_{41}|^2$ as a function of the side-gate voltage $u_g$, while fixing the remaining parameters as $\gamma_{sp}=0.4$, $\gamma_{sf}=0.1$, $\rho_{0}u=3$. Thus we expect that, under Coulomb blockade regime ($u_{g}=-3$), the non-local correlations $\mathcal{S}^{(th)}_{ij}/S_{th}^{U}$ present a maximum or a minimum depending on the couple of terminals considered. This behavior is directly related to the scattering amplitudes dependence on the side-gate voltage. Indeed, the amplitude $|S_{41}|^2$ is of order of one far from the Coulomb blockade region, while it is reduced when the Coulomb blockade is reached. Under the latter condition, the amplitudes $|S_{31}|^2 \sim \gamma_{sf}^2$ and $|S_{21}|^2 \sim \gamma_{sp}^2$ are fed in order to preserve the charge conservation law. The dependence of $\mathcal{S}^{(th)}_{31}/S_{th}^{U}=-|S_{31}|^2$ on the spin-flipping tunneling rate $\gamma_{sf}$ is shown in the lower panel of Fig.~\ref{fig:fig6}. In that figure, the $\mathcal{S}^{(th)}_{31}/S_{th}^{U}$ \textit{vs} $u_g$ curves present a minimum for $u_{g}=-3$ (Coulomb Blockade regime)  which becomes deeper when the value of $\gamma_{sf}$ is increased. Since $\mathcal{S}^{(th)}_{31}=0$ for $\gamma_{sf}=0$, non-local correlations provide a direct measure of the spin-orbit interaction reactivation in the QPC region.\\
Thus an accurate analysis of the thermal fluctuations allows a complete characterization of the junction parameters and put in evidence the helical nature of the system.

\subsection{Transport fluctuations}
\label{ssec:transport-fluct}
\begin{figure}[!h]
\includegraphics[scale=0.6]{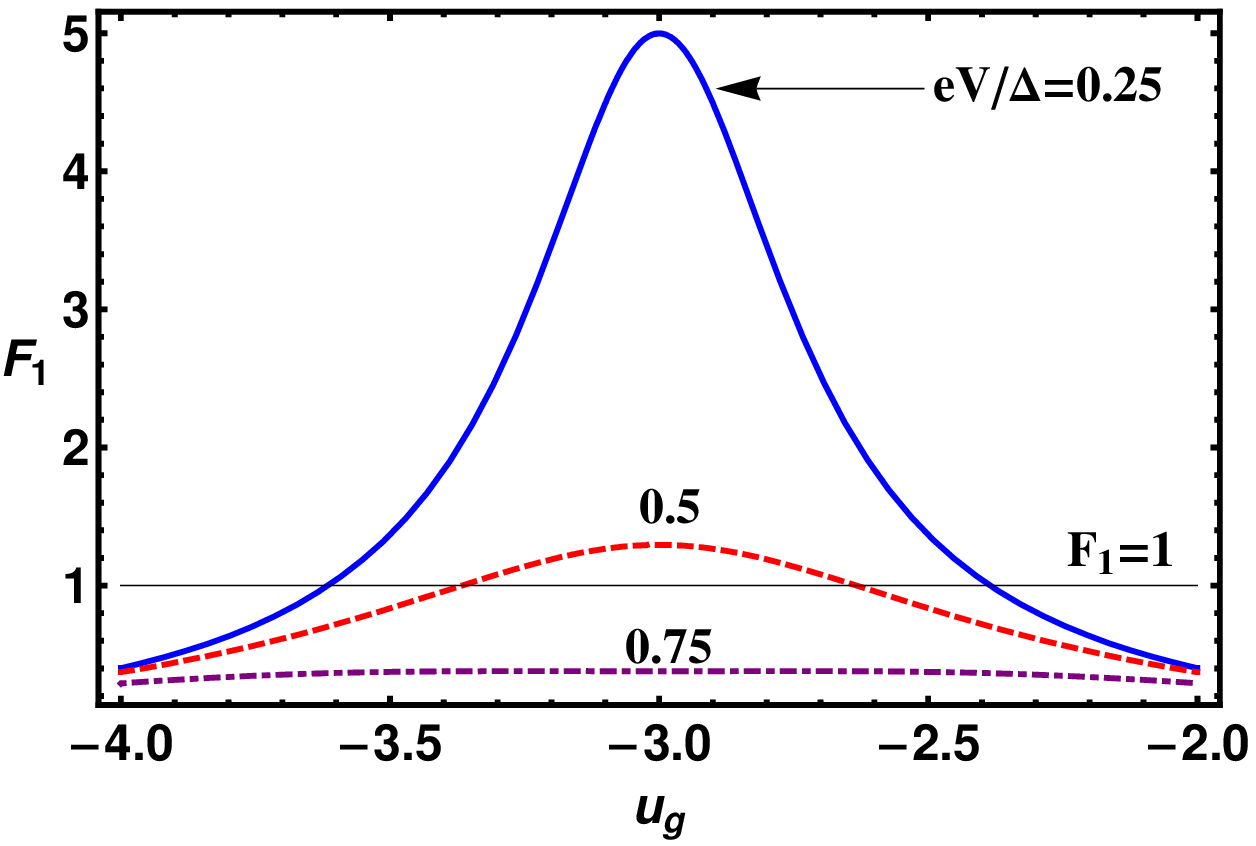}\\
\includegraphics[scale=0.6]{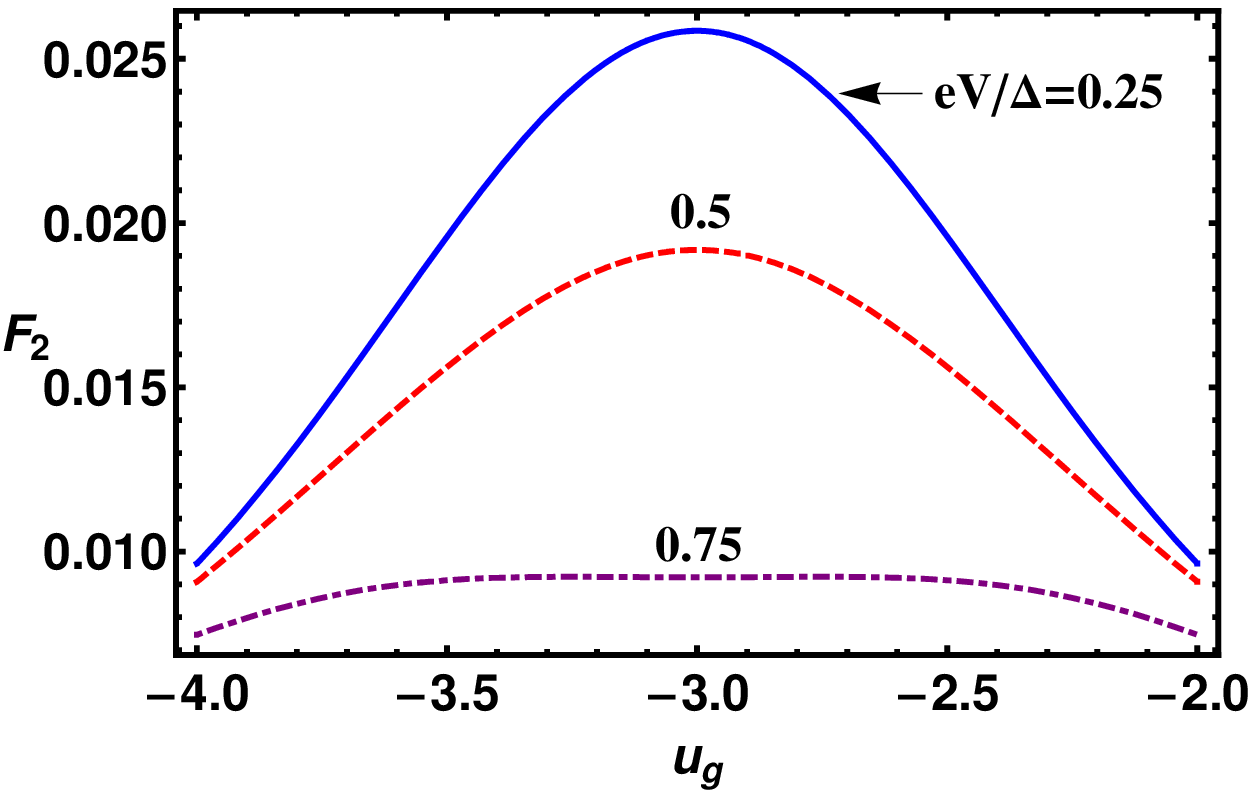}
\caption{(color online) Upper panel: $F_{1}$ \textit{vs} $u_{g}$ curves computed assuming junction parameters: $\gamma_{sp}=0.4$, $\gamma_{sf}=0.1$, $\rho_{0}u=3$ (as fixed in Fig.~3). The horizontal line $F_{1}=1$ is a guide for the eyes indicating the Fano factor of a Poissonian process for comparison. Lower panel: $F_{2}$ \textit{vs} $u_{g}$ computed by setting the junction parameters as in the upper panel. The different curves in the upper and lower panel refer to different values of applied voltage: $eV/\Delta=0.25$, $0.5$, $0.75$.}
\label{fig:fig7}
\end{figure}

We now consider current fluctuations in the presence of a steady state current and assume the zero-temperature limit. This regime is suitable for characterizing the non-equilibrium topological phase at the mK temperature range.
Under these assumptions and considering the low-frequency regime $\Omega \rightarrow 0$, the diagonal terms $\mathcal{F}_{ii}(0)\propto k_{B}T$ of the integral function in Eq.~(\ref{eq:spectral}) are negligible and thus the transport fluctuations can be written as
\begin{equation}
\label{eq:spectral_tr}
\mathcal{S}^{(tr)}_{ij} \equiv \mathcal{S}_{ij}(\Omega=0)=\frac{e^2}{h}\sum_{mn}S^{\ast}_{im}S_{jm}S^{\ast}_{jn}S_{in}\mathcal{F}_{mn}(0),
\end{equation}
where $\mathcal{F}_{mn}(0)$ is well approximated by the expression
\begin{equation}
\mathcal{F}_{mn}(0)=|e|\Bigl[-V_{m}-V_{n}-2~\mathrm{min}(-V_{m},-V_{n}) \Bigl].
\end{equation}
Once again we recall that Eq.~(\ref{eq:spectral_tr}) is non-linear with respect to the applied bias due to the hidden bias dependence of the scattering matrix, while the low-frequency correlations sum rule $\sum_{ij} \mathcal{S}^{(tr)}_{ij}=0$ has been numerically verified with accuracy within the numerical error ($\sim 10^{-16}$ in dimensionless units). The statistical properties of current fluctuations in the ith lead can be described by introducing the Fano factor $F_{i}=\mathcal{S}_{ii}^{(tr)}/(2|e|| I_{ch}^{(i)}|)$. Fano factor of 1 indicates Poissonian fluctuation processes, while sub-Poissonian (super-Poissonian) fluctuations are characterized by $F_{i}<1$ ($F_{i}>1$). Within the crossed bias configuration (i.e. $V_{1}=V_{3}=0$, $V_{2}=V/2$ and $V_{4}=-V/2$) here considered, introducing the notation $T_{ij}=|S_{ij}|^2$, one can explicitly derive the Fano factors which have the following expressions:
\begin{eqnarray}
F_{1}&=&\frac{T_{12}(1-T_{12})+T_{14}(1-T_{14})}{T_{14}-T_{12}}\nonumber\\
F_{2}&=&\frac{T_{24}(1-T_{24})}{1+T_{24}}.
\end{eqnarray}
They depend on the applied bias $V$ and on the side-gate voltage $u_g$ via the scattering amplitudes $T_{ij}$. In Fig.~\ref{fig:fig7}
we plot the Fano factor characterizing the fluctuation processes of the lead 1 (upper panel) and 2 (lower panel) as a function of the
side-gate voltage $u_{g}$ and by fixing the junction parameters as done in Fig.~\ref{fig:fig3}. The different curves both in the upper and lower
panel are computed taking bias voltages $eV/\Delta=0.25$, $0.5$, $0.75$, respectively.
The upper panel of Fig.~\ref{fig:fig7} shows that by tuning the side-gate voltage within the Coulomb blockade region a super-Poissonian noise is observed with a strong enhancement of the Fano factor up to a value of 5. Sub-Poissonian fluctuations are instead detected outside the blockade region. The Coulomb blockade regime can be overcome by using the side-gate voltage or by increasing the applied voltage bias. The Fano factor $F_{2}$ associated to the fluctuation in the lead 2 (see lower panel) is instead quite insensitive to the effect of $u_{g}$ which only produces a moderate increase of a sub-Poissonian Fano factor when the Coulomb blockade region is reached.
The occurrence of super-Poissonian noise can be ascribed to different mechanisms even though, in all the cases, the Coulomb blockade physics seems to be fundamental for the shot noise enhancement\cite{zhang2007,zarchin2007,ubbelohde2013,onac2006,safonov2003}. In the lead 1 two conditions favorable to the shot-noise enhancement are verified: (i) two channels, with different transparencies, are available for transport, i.e. the paths linking the couple of electrodes $1-2$ and $1-4$; (ii) the Coulomb blockade regime makes the tunneling events correlated and thus the electrons are transferred in bunches when transport takes place through the more transparent channel.\\
On the other hand, the charge current in the lead 2 presents an ohmic behavior coming from the interaction of an open channel (carrying a current $\frac{e^2}{h}\frac{V}{2}$) with a low-transparency channel (carrying a current $\frac{e^2}{h}\frac{V}{2}T_{24}$). The open channel does not contribute to the current fluctuations, while the opaque one is responsible for the sub-Poissonian value of $F_{2}$ since $T_{24} \sim \gamma_{sf}^2 \ll 1$.\\
From the experimental side, it has been reported in Ref.~[\onlinecite{safonov2003}] that the presence of coexisting current paths, namely a hopping (diffusive) path in parallel with a resonant tunneling process, hampers the manifestation of an enhanced Fano factor. In that case, the measured Fano factor $F$ results from a weighted average of the diffusive Fano factor $F_{B} \sim 0.33$, associated to the hopping background, and of the Fano factor $F_{RT}$ describing the resonant tunneling process ($F_{RT} \approx 8$). When the diffusive and the resonant channels contribute equally to the current, the measured Fano factor $F$ is near equal to $F_{RT}/2$, while a further reduction of $F$ is expected if the diffusive path dominates the transport. The latter situation, however, appears favored in devices with large transverse dimension ($W \sim 20~\mu m$ in the case of Ref.~[\onlinecite{safonov2003}]) where the presence of diffusive channels alternative to the direct tunneling transmission cannot be excluded.
On the other hand, the possibility to observe effective Fano factors of order 10 is supported by Ref.~[\onlinecite{zarchin2007}] where edge-channel transport with quantized conductance $\nu e^2/h$ has been induced by using high magnetic field in the range 9-11 T.
Thus, in our opinion, the Fano factor enhancement here reported can in principle be detected in a topological corner junction where the presence of diffusive effects should be limited by the insulating bulk of the topological insulator. Within the QPC region, however, the presence of impurity states coupled to the edge-channels can provide diffusive links which lower the effective Fano factor. For the above reasons, experimental investigations are required in order to clarify the role of the topological protection in promoting super-Poissonian fluctuations with giant Fano factor in Coulomb blockade regime.

\subsection{Non-local Transport fluctuations}
\label{ssec:non-local-transport-fluct}
\begin{figure*}[!t]
\includegraphics[scale=0.5]{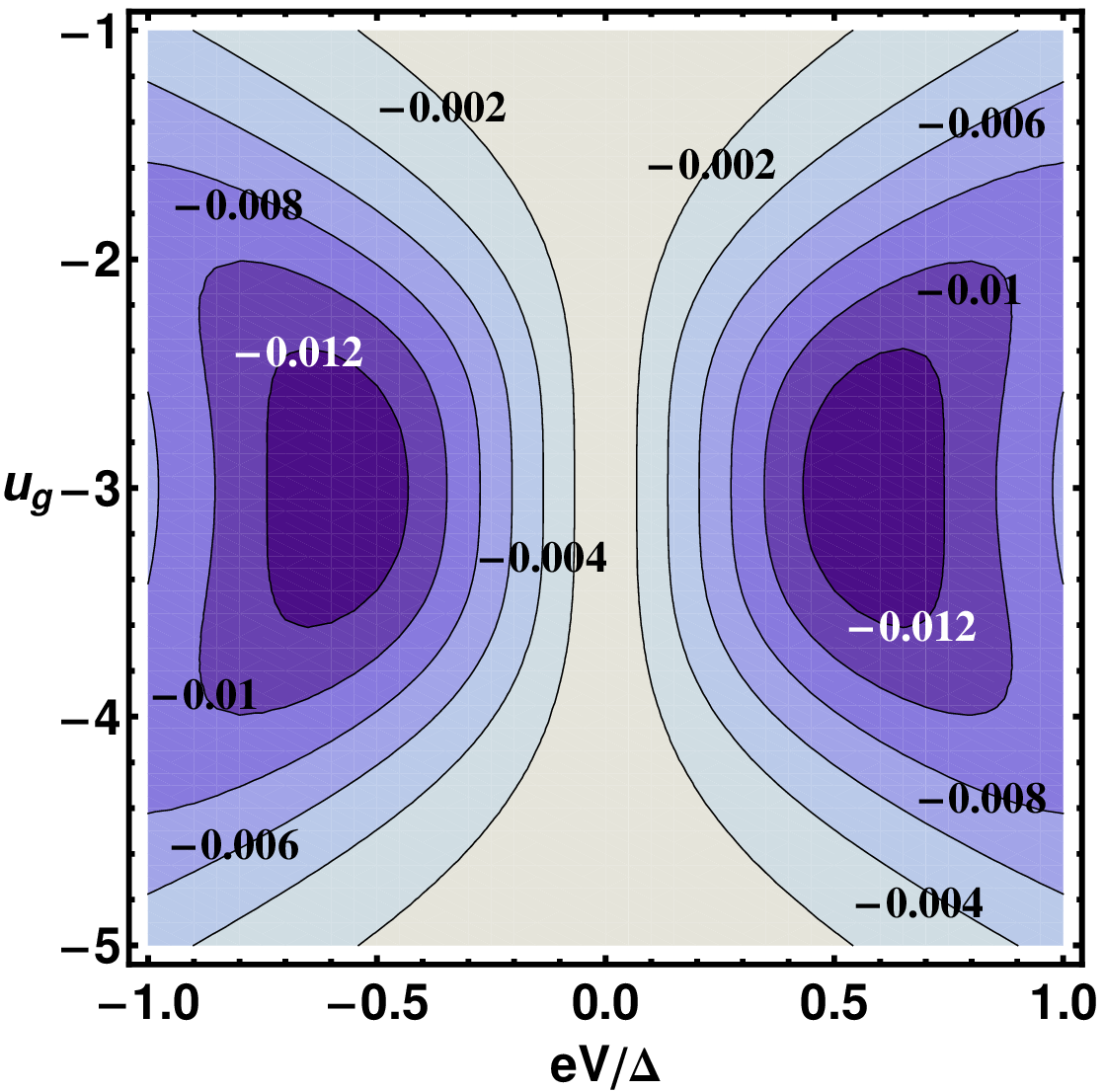} \includegraphics[scale=0.5]{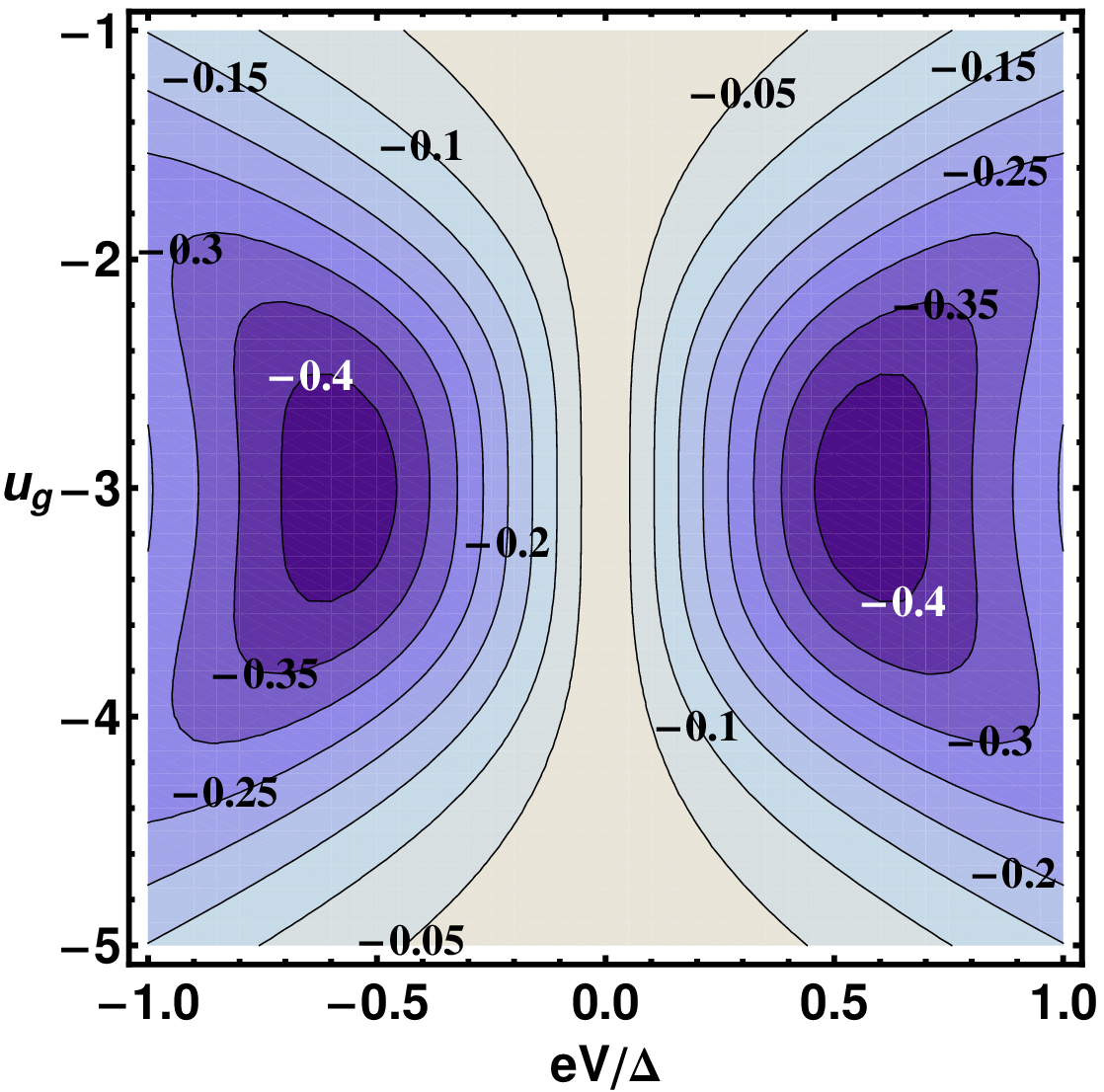} \includegraphics[scale=0.5]{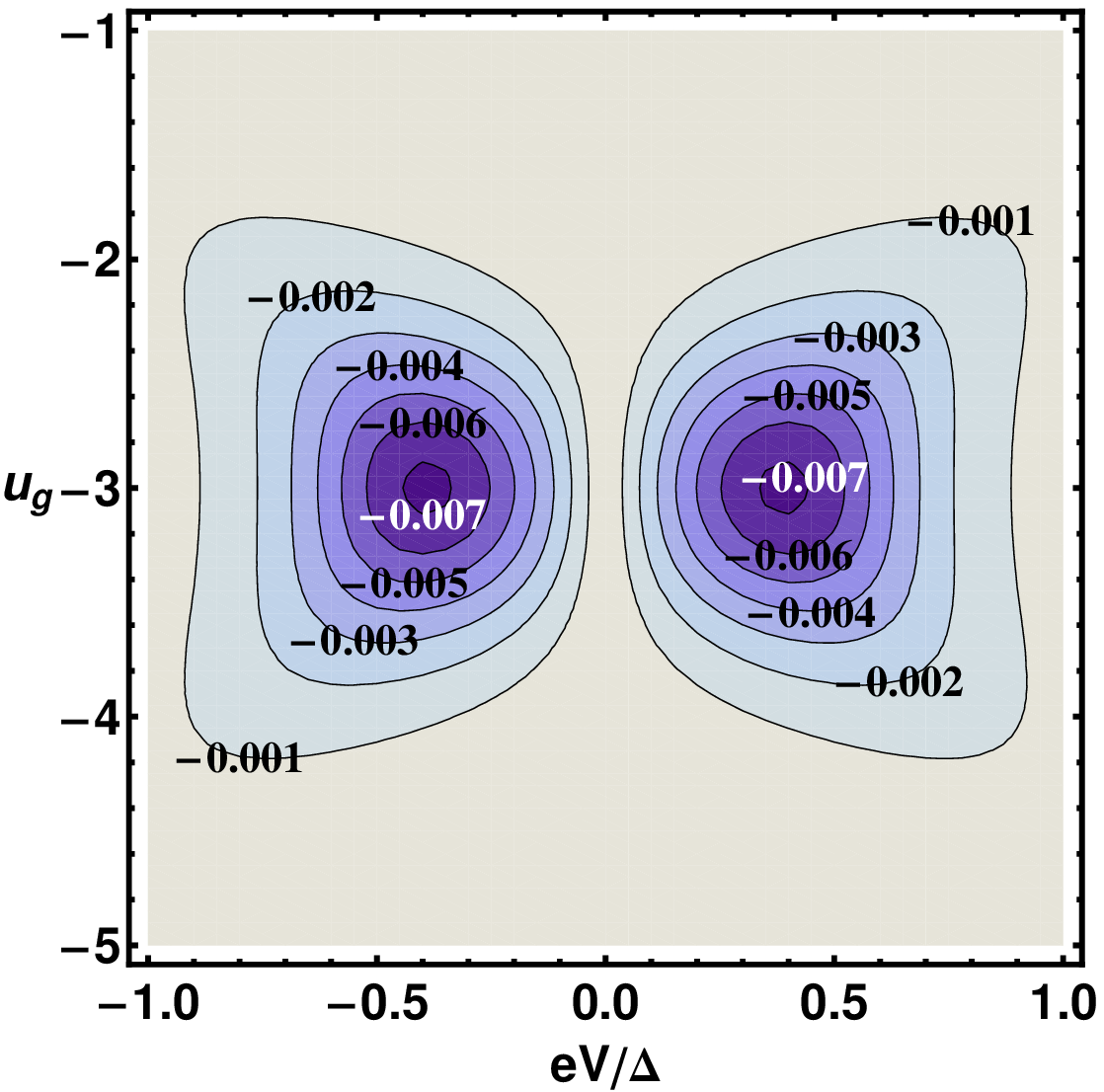}
\caption{(color online) Correlation functions (in unit of $e^{2}\Delta_{bulk}/(2h)$) $\mathcal{S}^{(tr)}_{12}$ (left panel), $\mathcal{S}^{(tr)}_{13}$ (middle panel), and $\mathcal{S}^{(tr)}_{14}$ (right panel) as a function of the applied bias $eV/\Delta$ and of the side-gate voltage $u_g$. The junction parameters are fixed as follows: $\gamma_{sp}=0.4$, $\gamma_{sf}=0.1$ and $\rho_{0}u=3$.}
\label{fig:fig8}
\end{figure*}

Here we report the analysis of the non-local current correlations assuming the zero temperature limit. Starting from Eq.~(\ref{eq:spectral_tr}), we derive the expressions of $\mathcal{S}^{(tr)}_{12}$, $\mathcal{S}^{(tr)}_{13}$ and $\mathcal{S}^{(tr)}_{14}$ as a function of the scattering matrix elements:
\begin{eqnarray}
\label{eq:non-loc-corr}
\mathcal{S}^{(tr)}_{12}&=&\frac{e^{3}|V|}{h}\mathrm{Re}\Bigl\{S_{13}S^{\ast}_{23}S^{\ast}_{14}S_{24} \Bigl\}\nonumber\\
\mathcal{S}^{(tr)}_{13}&=&2\frac{e^{3}|V|}{h}\mathrm{Re}\Bigl\{ S_{14}S^{\ast}_{34}S^{\ast}_{12}S_{32}\Bigl\}\nonumber\\
\mathcal{S}^{(tr)}_{14}&=&\frac{e^{3}|V|}{h}\mathrm{Re}\Bigl\{S_{13}S^{\ast}_{43}S^{\ast}_{12}S_{42} \Bigl\}.
\end{eqnarray}
These correlation functions present non-linear dependence on the applied bias $V$ (encoded within the scattering matrix elements) and contain information on the tunneling amplitude renormalization in Coulomb blockade regime.
In the absence of charging effects and for $u_{g}=0$ ($u_{1}=u_{2}=0$) Eqs.~(\ref{eq:non-loc-corr}) can be written in closed form in terms of the bare junction parameters $\gamma_{sp}$ and $\gamma_{sf}$ according to the following expressions:
\begin{eqnarray}
\label{eq:non-loc-corr-free}
\mathcal{S}^{(tr)}_{12}&=&-4\frac{e^{3}|V|}{h}\frac{(1-\Gamma^2)^2 \gamma_{sf}^2}{(1+\Gamma^2)^4}\nonumber\\
\mathcal{S}^{(tr)}_{13}&=&-8\frac{e^{3}|V|}{h}\frac{(1-\Gamma^2)^2 \gamma_{sp}^2}{(1+\Gamma^2)^4}\nonumber\\
\mathcal{S}^{(tr)}_{14}&=&-16\frac{e^{3}|V|}{h}\frac{(\gamma_{sp}\gamma_{sf})^2}{(1+\Gamma^2)^4},
\end{eqnarray}
where the notation $\Gamma^2=\gamma_{sp}^2+\gamma_{sf}^2$ has been introduced. Eqs.~(\ref{eq:non-loc-corr-free}) show that in the absence of spin-flipping tunneling, i.e. for $\gamma_{sf}=0$, the only non-vanishing correlation is $\mathcal{S}^{(tr)}_{13}$, being $\mathcal{S}^{(tr)}_{12}=\mathcal{S}^{(tr)}_{14}=0$. Since the above observation remains true also in the case of non-vanishing Coulomb interaction, studying the non-local current correlations provides the opportunity to further characterize the junction and, in particular, allows the detection of the spin-orbit interaction reactivation in the constriction region. For corner junctions characterized by a moderate reactivation of the spin-orbit effect ($\gamma_{sp}>\gamma_{sf}$), since $\mathcal{S}^{(tr)}_{12} \sim \gamma_{sf}^2$, $\mathcal{S}^{(tr)}_{13}\sim \gamma_{sp}^2$ and $\mathcal{S}^{(tr)}_{14}\sim (\gamma_{sp}\gamma_{sf})^2$, the above equations predict the following relation among the strengths of the correlation functions: $|\mathcal{S}^{(tr)}_{13}|>|\mathcal{S}^{(tr)}_{12}|>|\mathcal{S}^{(tr)}_{14}|$.\\
Moreover the non-local correlations contain information on the Coulomb blockade regime and on the tunneling amplitudes renormalization due to the interaction. This analysis is performed in Fig.~\ref{fig:fig8} where the correlation functions (in unit of $e^{2}\Delta_{bulk}/(2h)$) $\mathcal{S}^{(tr)}_{12}$, $\mathcal{S}^{(tr)}_{13}$, and $\mathcal{S}^{(tr)}_{14}$ are shown as a function of the applied bias $eV/\Delta$ and of the side-gate voltage $u_g$.
In particular, all the correlation functions show negative values and present a double-minimum structure within the Coulomb blockade region. The two minima are located at $eV/\Delta \approx  \pm 0.5$ and $u_{g}=-3$, the latter bias configuration defining the edge of the Coulomb blockade region. Thus, the absolute value of the non-local correlations is maximized when the Coulomb blockade regime is going to be overcome (see Fig.~\ref{fig:fig2}). Interestingly, the presence of the two minima in the correlation functions is directly related to the interaction and its renormalization effect on the tunneling amplitudes. This is supported by the fact that, in the absence of charging effects, the correlation functions present a bias dependence proportional to $-|V|$ [see Eqs.~(\ref{eq:non-loc-corr-free})] which does not allow the presence of minima. Moreover, Figs.~\ref{fig:fig8} confirm the relation $|\mathcal{S}^{(tr)}_{13}|>|\mathcal{S}^{(tr)}_{12}|>|\mathcal{S}^{(tr)}_{14}|$ which maintains its validity also in the presence of charging effects.
\section{Conclusions}
\label{sec:concl}
We theoretically studied the transport properties of a corner-junction realized by using a two-dimensional topological insulator shaped in the
form of a four-terminal QPC. The Coulomb interaction effects, at mean field level, and the reactivation of the spin-orbit coupling due to the tight confinement of the electron densities within the QPC region are taken into account by implementing a self-consistent scattering field theory able to capture the Coulomb blockade physics. In this framework, the charging effects are described by a low-energy scattering matrix parametrized by two self-consistent parameters, namely $u_{1}$ and $u_{2}$, and the resulting theoretical approach takes simultaneously into account non-equilibrium effects and Coulomb blockade physics beyond the linear response regime. Assuming a crossed-bias configuration ($V_{1}=V_{3}=0$ and $V_{2}=-V_{4}=V/2$), we studied the current-voltage characteristics of the device and the current-current local and non-local correlations (noise). The dc transport of the device shows a Coulomb blockade regime which can be removed by using the side-gate and the bias voltage, while the differential conductance presents a structure reminiscent of a single Coulomb diamond whose extension along the side-gate axis is proportional to the interaction value $\rho_0 u$. Concerning the fluctuation properties of the current flowing through the device, we studied the thermal noise and the transport noise. It has been proven that the (local) thermal noise takes the universal value $S_{th}^{U}=4k_BT(e^2/h)$ below a gap-dependent characteristic temperature $T^{\ast}$, the latter property being relevant in discriminating devices with helical properties. The non-local thermal fluctuations contain instead important information on the tunneling amplitudes and thus can help in characterizing the junction properties. The Coulomb blockade region has been further characterized  by evaluating the Fano factor which provides a measure of the transport fluctuations in a given electrode. It has been found a sub-Poissonian character of the fluctuation outside the Coulomb blockade region and a super-Poissonian value of the Fano factor as the effect of the blockade imposed by the charging energy within the QPC. The non-local transport fluctuations exhibit a double-minimum structure as a function of the applied bias. The double-minimum is peculiar of a non-vanishing Coulomb interaction within the QPC and thus is not present when the charging energy of the constriction is neglected.
The theory presented in this work contains all the relevant ingredients needed to describe the transport properties of a corner-junction realized using helical matter (i.e. a topological insulator) and allows a direct comparison with forthcoming experimental works. The experimental implementation of our proposal presents similar difficulties of the ones described in Ref.~[\onlinecite{zarchin2007}] even though the topological corner-junction setup does not require the use of high magnetic fields to induce edge states, the latter being an important simplification of the experimental situation. Finally, it's worth mentioning that the universal nature of the thermal fluctuation of a corner-junction realized using helical matter can be an important signature to probe the emergence of a topological phase at cryogenic temperatures (few Kelvins). The latter point appears to be even more important in testing topological properties of the next-generation room temperature topological insulators\cite{kuo13}.

\appendix

\begin{widetext}
\section{$\mathcal{M}$ Matrix}
\label{app:A}
The explicit expression of the $\mathcal{M}$ matrix in terms of the scattering matrix elements takes the following form:

\begin{equation}\label{eq:m}
\mathcal{M}=-\frac{u\rho_{0}}{4}\left(
                                  \begin{array}{cccc}
                                    |S_{21}|^2+ |S_{31}|^2& 1+ |S_{32}|^2& 1+ |S_{23}|^2& |S_{24}|^2+ |S_{34}|^2 \\
                                    1+ |S_{41}|^2 & |S_{12}|^2+ |S_{42}|^2 & |S_{13}|^2+ |S_{43}|^2 & 1+ |S_{14}|^2 \\
                                  \end{array}
                                \right).
\end{equation}

\section{Scattering matrix }
\label{app:B}
The explicit expression of the scattering matrix $S[u_{1},u_{2},\gamma_{sp},\gamma_{sf}]$ in terms of the bare tunneling amplitudes $\gamma_{sp/sf}$ and of the self-consistency potentials $u_{1/2}$ takes the following form:
\begin{equation}\label{eq:sm}
\left(
                 \begin{array}{cccc}
                   0 & -\frac{2i \gamma_{sp}}{1+i(u_{1}+u_{2})-u_{1}u_{2}+\gamma_{sp}^2+\gamma_{sf}^2} & \frac{2i \gamma_{sf}}{1+i(u_{1}+u_{2})-u_{1}u_{2}+\gamma_{sp}^2+\gamma_{sf}^2} & \frac{1+i(u_{1}-u_{2})+u_{1}u_{2}-\gamma_{sp}^2-\gamma_{sf}^2}{1+i(u_{1}+u_{2})-u_{1}u_{2}+\gamma_{sp}^2+\gamma_{sf}^2} \\
                   -\frac{2i \gamma_{sp}}{1+i(u_{1}+u_{2})-u_{1}u_{2}+\gamma_{sp}^2+\gamma_{sf}^2}& 0 & \frac{1-i(u_{1}-u_{2})+u_{1}u_{2}-\gamma_{sp}^2-\gamma_{sf}^2}{1+i(u_{1}+u_{2})-u_{1}u_{2}+\gamma_{sp}^2+\gamma_{sf}^2} & \frac{2i \gamma_{sf}}{1+i(u_{1}+u_{2})-u_{1}u_{2}+\gamma_{sp}^2+\gamma_{sf}^2} \\
                   -\frac{2i \gamma_{sf}}{1+i(u_{1}+u_{2})-u_{1}u_{2}+\gamma_{sp}^2+\gamma_{sf}^2} & \frac{1-i(u_{1}-u_{2})+u_{1}u_{2}-\gamma_{sp}^2-\gamma_{sf}^2}{1+i(u_{1}+u_{2})-u_{1}u_{2}+\gamma_{sp}^2+\gamma_{sf}^2} & 0 & -\frac{2i \gamma_{sp}}{1+i(u_{1}+u_{2})-u_{1}u_{2}+\gamma_{sp}^2+\gamma_{sf}^2} \\
                   \frac{1+i(u_{1}-u_{2})+u_{1}u_{2}-\gamma_{sp}^2-\gamma_{sf}^2}{1+i(u_{1}+u_{2})-u_{1}u_{2}+\gamma_{sp}^2+\gamma_{sf}^2} & -\frac{2i \gamma_{sf}}{1+i(u_{1}+u_{2})-u_{1}u_{2}+\gamma_{sp}^2+\gamma_{sf}^2} & -\frac{2i \gamma_{sp}}{1+i(u_{1}+u_{2})-u_{1}u_{2}+\gamma_{sp}^2+\gamma_{sf}^2} & 0 \\
                 \end{array}
               \right).
\end{equation}
\end{widetext}

\section*{Acknowledgements}
The authors gratefully acknowledge D. Ferraro and A. Braggio for helpful comments. Discussions with C. Barone and S. Pagano on the 1/f noise in nanodevices are also acknowledged. One of the authors (R. C.) acknowledges support from the Project FIRB-2012-HybridNanoDev.


\begin{thebibliography}{99}
\bibitem{note1} Where not specified the integration domain is given by $[-\infty,+\infty]$, i.e. $\int dx \equiv \int_{-\infty}^{\infty} dx$.
\bibitem{nota_int_fermi} At low temperature we have:
\begin{equation}
\int_{\frac{-\Delta_{bulk}}{2}}^{\frac{\Delta_{bulk}}{2}} f_{i}(E) dE \approx \mu-|e|V_{i}+\frac{\Delta_{bulk}}{2}\nonumber,
\end{equation}
where the insulating bulk gap energy $\Delta_{bulk}$ represents the energy cut-off of the edge-states theory.
\bibitem{note2} For the grounded configuration ($V_{j}=0$, $\forall j$) the self-consistent equations given in Eqs.~(\ref{eq:self2}) are directly solved by $u_{1}=u_{2}=\rho_{0}u+u_{g}$. Thus the scattering matrix of the problem is given in closed form by Eq.~(\ref{eq:sm}) with the substitution $u_{1}=u_{2}=\rho_{0}u+u_{g}$. Moreover, in this case, Coulomb interaction just redefines the zero of the side-gate voltage $u_{g}$.
\end{thebibliography}
\end{document}